\documentclass[showpacs,aps,prd,twocolumn,eqsecnum,floats,superscriptaddress,%
nofootinbib,floatfix,preprintnumbers]{revtex4}

\usepackage{graphicx}
\usepackage{amsfonts}
\usepackage{amsmath}
\usepackage{subfigure}
\usepackage{bm}
\usepackage{epsfig}

\def\msun{\mathrm{M}_\odot}

\def\htilde{\tilde{h}(f)}
\def \st{\tilde{s}}

\def \A {{\cal A}}

\def \Q{{\vec Q}}
\def \H {{\cal H}}

\def \Nind {N_{\rm ind}}
\def \Nwin {N_{\rm win}}

\def \e0{\epsilon_0}
\def \W {{\cal W}}
\def \h {{1 \over 2}}

\def \DW {\Delta_{\W}}
\def \Dn {\Delta_n}

\def\be{\begin{equation}}
\def\ee{\end{equation}}
\def\bea{\begin{eqnarray}}
\def\eea{\end{eqnarray}}
\def \no {\nonumber}
\def\lsim{\mathrel{\rlap{\lower4pt\hbox{\hskip1pt$\sim$}}
    \raise1pt\hbox{$<$}}}                
\def\gsim{\mathrel{\rlap{\lower4pt\hbox{\hskip1pt$\sim$}}
    \raise1pt\hbox{$>$}}}                

\begin{document}

\preprint{IUCAA 38/09\cr
OU-TAP 300}

\title{Detecting gravitational waves from inspiraling binaries with a network of geographically separated detectors : coherent versus coincident strategies}

\author{Himan Mukhopadhyay}
\affiliation{Inter-University Centre for Astronomy and Astrophysics,\\
Post Bag 4, Ganeshkhind, Pune 411007, India}

\author{Hideyuki Tagoshi} 
\affiliation{Department of Earth and Space Science, 
Graduate School of Science, Osaka University, Toyonaka, 
Osaka 560-0043, Japan}

\author{Sanjeev Dhurandhar}
\affiliation{Inter-University Centre for Astronomy and Astrophysics,\\
Post Bag 4, Ganeshkhind, Pune 411007, India}

\author{Nobuyuki Kanda}
\affiliation{Department of Physics, Graduate School of Science, Osaka City University, 
Osaka 558-8585, Japan}


\begin{abstract}
We compare two strategies of multi-detector detection of compact binary inspiral signals, namely, the coincidence and the coherent for the realistic case of geographically separated detectors. The {\it naive} coincident strategy treats the detectors as if they are isolated - compares individual detector statistics with their respective  thresholds while the coherent strategy combines the detector network data {\it coherently} to obtain a single detection statistic which is then compared with a single threshold. We also consider an {\it enhanced} coincidence strategy which is intermediate in the sense that though the individual statistics are added in quadrature and the sum compared with a single threshold, the estimated parameters are also checked for consistency. For simplicity, we consider detector pairs having the same power spectral density of noise, as that of initial LIGO and also assume the noise to be stationary and Gaussian. Further, since we consider the detectors to be widely separated on Earth, we take  the instrumental noises to be uncorrelated; the wide separation implicitly means that since the detector arms must lie parallel to the Earth's surface, the detectors necessarily have different orientations.  We compare the performances of the  methods by plotting the \emph{receiver operating characteristics} (ROC) for the  strategies.  Several results are derived analytically in order to gain insight. Simulations are performed in order to plot the ROC curves. A single astrophysical source as well as a distribution of sources is considered. We find that the coherent strategy is superior to the two coincident strategies that we consider. Remarkably, the detection probability of the coherent strategy is 50\% better than the naive coincident strategy. One the other hand, difference in performance  between the coherent strategy and enhanced coincident strategy is not very large.
Even in this situation, it is not difficult to perform the real data analysis with the coherent strategy. The bottom line is that the coherent strategy is a good detection strategy.
\end{abstract}
\maketitle

\section{Introduction}

Inspiraling binaries are one of the most promising candidates for first detection of gravitational waves (GW). The compact objects can be treated essentially as point particles  leading to sufficiently adequate description of the system in terms of the post-Newtonian formalism \cite{blanchet}. The great accuracy of the post Newtonian approximation of the phase, about a cycle for a wave train $\sim 10^4$ cycles long, renders it amenable for matched filtering analysis \cite{arun}. Inspiraling binaries are  astrophysically important, because they will not only carry detailed information about the binary system, but also general relativistic deviations from Newtonian gravity in their orbit can  experimentally be measured \cite{binary1, binary2}. The best available estimates suggest that at $1\%$ false alarm probability the expected number of neutron star(NS)-NS binary coalescence seen per year by ground based interferometers is $3 \times 10^{-4} - 0.3$ for initial detectors and $1-800$ for advanced detectors \cite{kalogera}. In recent years, a number of ground based detectors are taking quality science data and are collaborating together, thus the time is ripe to consider analysis of network data for the detection of inspiraling binaries. The advantages of multi-detector search for the binary inspiral is that, not only does it improve the confidence of detection, but it also provides information about the direction and polarization state of the source. 

Two strategies currently exist in searching for inspiraling binary sources with a network of detectors: the coherent and the coincident. The coherent strategy involves combining data from different detectors phase coherently, appropriately correcting for  time-delays and polarization phases and obtaining a single statistic for the full network, that is optimized in the maximum likelihood sense. On the other hand, the coincident strategy matches the candidate event lists of individual detectors for consistency of the estimated parameters of the GW signal. A coincidence search with real data has been carried out by several group \cite{pastcoin} and data analysis on S5 data of the LIGO detectors has recently been performed \cite{LIGOCBC}. 

There is a long standing debate as to which strategy performs better. In an earlier paper \cite{himan}, hereafter referred to as paper-I, we compared the performances for the simple case of coaligned detectors located in the same place. The situation of two coaligned detectors located at the same place with correlated noise was first considered in paper-I and improved in \cite{tagoshi}, is applicable in the case of existing detector pairs H1-H2. The coincident strategy has the advantage of reduction of false alarm, however this is at the price of reduced detection efficiency. On the other hand, in the coherent strategy, the false alarm rate is not reduced, the sensitivity is enhanced which in turn results in higher detection efficiency. Which of these two competing effects wins was determined by looking at the detection efficiencies of the two strategies at the same false alarm rate. \emph{Receiver operating characteristics} (ROC) curve, which is the plot of detection efficiency versus false alarm rate, was drawn for both the strategies and from those curves it was inferred that for the viable false alarm regime the coherent strategy performs much better than the coincident strategy.

In this paper we consider the general case of widely separated pairs of detectors. Since the detectors are situated on the globe and have their arms lying parallel to the Earth's surface, they necessarily have different orientations;  we however consider the general problem of detectors with arbitrary locations and orientations, since this does not greatly add to the mathematical complexity of the problem.  The coherent statistic for non-aligned detectors is completely different from that of the aligned case \cite{PDB}. We cannot extrapolate the results of paper-I to incorporate the more general case that we consider here. We compare the performance of the two strategies for the generalized case by  obtaining the relevant ROC curves. We then investigate an enhanced coincidence strategy which is basically an improvement on the simple coincidence strategy for it is seen that when the detectors are not aligned, the performance of the simple coincidence strategy is very poor. In enhanced coincidence, although the two detectors are considered in isolation and the candidate event lists  compared for consistency in the estimated signal parameters, the two statistics from individual detectors are added in quadrature and a single threshold is placed upon the resulting statistic. This strategy, for the case of two detectors only, renders a statistic identical in form to that of the coherent strategy. The details of this strategy are given in section III B.2. It should be noted that it is different from the coherent strategy.

The plan of the paper is as follows. In section \ref{signal}, we discuss the different coordinate systems required to describe the problem, the signal and the response of the detectors and the network. In section \ref{form}, we discuss the false alarm and detection probabilities and derive  the  analytical formulae for the coincidence and coherent strategies. In section \ref{comper} we perform simulations and then use these to plot the ROC curves for different detector pairs. We first briefly present the  simulation method and then we determine the parameter windows required for coincidence analysis. We perform simulations for arbitrarily oriented detector pairs and obtain the false alarm and detection probabilities for each pair and use these results to plot the ROC curves. In section \ref{conclusion}, we summarize our results and discuss future directions.

\section{The signal and the response}
\label{signal}

In this section we discuss the GW signal and the response of individual detectors as well as of the network and introduce a normalization scheme and notations.  We follow the conventions and notations of \cite{PDB}. In order to make the paper self contained we briefly review the formalism. It provides us with an efficient framework for the analysis here in the context of widely separated detectors with arbitrary orientations.

\subsection{Reference frames and Euler angles}
\label{frame}

To understand the response to the GW signal of the detectors and the network, the first step is to identify the different coordinate systems naturally associated with the problem and the interrelations between them. The frames are related via rotations described by Euler angles which then appear in the network response in the form of beam pattern functions. 

{\it Wave frame} - $(X,Y,Z)$ : The gravitational wave travels along the positive $Z$ direction and $X$ and $Y$ denote the axes of polarization - $(X,Y,Z)$ form a right-handed Cartesian coordinate system.

{\it Frame of detector $I$} - $(x_{I},y_{I},z_{I})$ : This is the frame of the detector $I$ having its origin at the intersection of the arms of the detector. The arms of the detector lie in the $x_{I}-y_{I}$ plane, which is the plane tangent to the surface of the Earth  with the $x_{I}$ axis bisecting the angle between the two arms and $y_{I}$ is chosen such that the frame forms a right-handed coordinate system with the $z_{I}$ axis pointing radially out of the surface of the Earth. Note that the Earth is assumed to be a sphere for the purposes of this analysis. 

{\it  Earth frame} - $(x,y,z)$: This is the Earth frame with its origin at the centre of the Earth. The $z$ axis points due north, the $x$ axis points to the intersection of the equator and the Greenwich meridian and the $y$ axis is chosen such that the coordinate system is right handed. This frame is usually chosen as the \emph{fiducial} frame of reference, with respect to which the orientations and locations of each detector can be specified.  In table \ref{orient} we specify the locations and orientations only of those detectors which we will use in our analysis.  This is clearly not an exhaustive list of ground-based detectors existing or planned and the results obtained in this paper can also be obtained similarly for other detectors.

\begin{table}[!hb]
\begin{tabular}{ccccc}
\hline
Detector & latitude($\lambda$) & longitude($\ell$) & x-arm($\zeta$) & y-arm \\
\hline
H1,H2 & $46.45^\circ$N & $-119.41^\circ$E & $36.8^\circ$ & $126.8^\circ$\\
L1 & $30.56^\circ$N & $-90.77^\circ$E & $108.0^\circ$ & $198.0^\circ$\\
V1 & $43.63^\circ$N & $10.5^\circ$E & $71.5^\circ$ & $341.5^\circ$\\
K1 & $ 137.18^\circ$N & $36.25^\circ$E & $295.0^\circ$ & $25.0^\circ$\\
\hline
\end{tabular}
\caption{The locations and orientations of some ground based GW detectors. 
L1 - LIGO Louisiana, H1 - LIGO Hanford 4 km, H2 - LIGO Hanford 2 km,
V1 - Virgo, and K1 - LCGT (the K stands for Kamioka).
The latitude and longitude of the centre and the angles through which the x-arm and the y-arm must be rotated clockwise (viewed from top) to point to North are listed. The angles $(\lambda, \ell, \zeta)$ are related to $(\alpha_I, \beta_I, \gamma_I)$ for a particular detector $I$.}
\label{orient}
\end{table}

Let $(\phi, \theta, \psi)$ be the Euler angles through which the Earth frame must be rotated in order to align with the wave frame. The angles $\phi$ and $\theta$ are closely related to spherical polar coordinates and $\psi$ is the polarization angle. We opt for the Goldstein convention \cite{Goldstein}. The Euler angles $(\alpha_I, \beta_I, \gamma_I)$ that rotate the Earth frame to the frame of the $I$-th detector are then $(\lambda_I+\pi/2, \ell_I, \zeta_I+3\pi/4)$.

\subsection{Signal at a detector}
\label{detectsignal}

Having described the coordinate system, we write down the response of the $I$-th detector to the GW signal in the Fourier domain. We adopt the following convention for the Fourier Transform,
\begin{equation}
\htilde = \int_{-\infty}^{\infty}dt\ h(t) e^{2\pi ift},
\label{eq:fftdef}
\end{equation}
as in paper-I.

In the stationary phase approximation the spin-less, restricted post-Newtonian 
inspiral signal is given by, 
\bea
\no
\tilde{h}^I(f) &=& {\mathcal{N}} \times E^I f^{-7/6} \\
&& \times \exp ~ i \Psi^I(f; t_c, \delta_c, \tau_0, \tau_3) ,
\label{eq:stilde}			   
\eea
where $t_c$ and $\delta_c$ are respectively the coalescence time and the coalescence phase of the binary. $E^I$ is the extended antenna pattern function which depends on seven angles - two angles $(\theta, \phi)$ specifying the location of the source, two angles $(\iota, \psi)$ specifying the orientation of the source and three Euler angles $(\alpha_I, \beta_I, \gamma_I)$ specifying the orientation of the detector with respect to the wave frame.
\be
E^I = \left ( \frac{1 + \cos^2 \iota}{2} \right ) F_+^{I} + i \cos \iota F_{\times}^{I} \,,
\label{exantptn} 
\ee
where the $F_{+, \times}^I$ are the usual antenna pattern functions depending on $\theta, \phi, \psi, \alpha_I, \beta_I, \gamma_I$ \cite{PDB}. Note that $E^I$ is a complex quantity. Thus the phase of $E^I$, say $ \chi^I$, where, $\tan \chi^I = 2 F^I_{\times} \cos \iota / F^I_{+} (1 + \cos^2 \iota )$ represents the polarization phase and plays an important role in coherent detection. On the other hand we observe that $\chi^I$ simply adds on to the coalescence phase $\delta_c$ and therefore for the purposes of matched filtering in a single detector can be absorbed in $\delta_c$. In the coincidence strategy each detector is treated separately, just like a single detector, and therefore, for matched filtering when using coincidence, $\chi^I$ can be absorbed in $\delta_c$.  

The factor $\mathcal{N}$ depends on the distance $r$ to the binaries, the total mass $M = m_1 + m_2$ and the reduced mass ratio $\eta = \frac{m_1 m_2}{M^2}$, where $m_1$ and $m_2$ are the individual masses of the binary. In the units of $c = G = 1$ it is given by, 
\be 
{\cal N} = \left(\frac{5}{24}\right)^{1/2}
\frac{M^{5/6}\eta^{1/2}\pi^{-2/3}}{r} \,. 
\label{N}
\ee 
Instead of the total mass and the chirp mass of the binary system, it is customary to use the dynamical parameters $\tau_0$ and $\tau_3$ for in these parameters the template placing is approximately uniform. These parameters are defined as: 
\bea
\no \tau_0 &=& \frac{5}{256 \pi \eta f_a} \left ( \pi M f_a \right )^{-5/3}, \\  
\tau_3 &=& \frac{1}{8 \eta f_a} \left ( \pi M f_a \right )^{-2/3},
\eea
where $f_a$ is the fiducial frequency usually chosen to be the seismic cut-off frequency - the lowest frequency of the detection bandwidth.

The phase of the signal $\Psi^I(f; t_c, \delta_c, \tau_0, \tau_3)$ in the $I$th detector frame relates to the phase in the fiducial frame $\Psi(f; t_c, \delta_c, \tau_0, \tau_3)$ as, 
\be
\Psi^I(f; t_c, \delta_c, \tau_0, \tau_3) = \Psi(f; t_c, \delta_c, \tau_0, \tau_3) + 2 \pi f \Delta t^I,
\ee
where $\Delta t^I$ is the time-delay between the $I$-th detector and the fiducial detector. The fiducial detector may be taken to coincide with the origin of the Earth frame, that is, a detector situated at the centre of the Earth or it could be chosen as one of the detectors in the network. For the phase in the fiducial frame  we adopt the same 3PN formula we took in paper-I (equations (2.7) and (2.8) in paper-I) given by \cite{arun}.

\subsection{The matched filtering paradigm for a network of detectors}
\label{nwsignal}

By virtue of (\ref{eq:stilde}), we can represent the $i$-th template for the $I$-th detector as, 
\bea
\no
\tilde{h}^I(f;\vec \mu_i, t_c, \delta_c) &=& \A^I (\st_0^I (f; \vec \mu_i, t_c) \cos \delta_c \\    
                                    &+&  \st_{\pi/2} ^I (f; \vec \mu_i, t_c) \sin \delta_c) \, ,
\label{eq:quadrature}
\eea
where,
\be
\st_{\pi/2}^I (f; \vec \mu_i, t_c) = i \st_{0}^I (f; \vec \mu_i, t_c) \,.
\label{eq:sopi} 
\ee
We require that both the templates $s_0^I$ and $s_{\pi/2}^I$ have have unit norm; i.e. the scalar products $(s_0^I,s_0^I)_I = (s_{\pi/2}^I, s_{\pi/2}^I)_I  = 1$. The scalar product $(a,b)_I$ of two real functions $a(t)$ and $b(t)$ for the $I$-th detector is defined as, 
\be
\left ( a, b \right )_I = 2 \int_{f_l^I}^{f_u^I} df \ \frac{  \tilde{a}(f) \tilde{b}^*(f) \ + \ 
\tilde{a}^*(f) \tilde{b}(f)}{S_h^I(f)},
\label{eq:scalar}
\ee
where, we use the Hermitian property of Fourier transforms of real functions. $S_h^I(f)$ is the one sided power spectral density (PSD) of the noise. Although in this paper we do not consider the case of different PSDs for different detectors, we still keep the discussion general so that it could also be used for the general case of different PSDs. This does not complicate the discussion much. 
It also follows from this definition that when the templates are computed using the stationary approximation, $(s_0^I, s_{\pi/2}^I)=0$, i.e., they are orthonormal.
Then, it is evident that $\A^I$ is the amplitude of the waveform, i.e., $(h^I,h^I)=(\A^I)^2$.  $f_l^I$ is the lower cut-off frequency which is normally taken to be the seismic cut-off -  40 Hz for initial LIGO - and $f_u^I$ is the the upper cut-off frequency usually taken to be about 1 kHz when the signal power for inspirals within the usual mass range, say, $1 M_{\odot} \leq m_1, m_2  \leq 40 M_{\odot}$ and initial LIGO PSD, falls of below a fraction of a percent. In most of the cases we have taken the noise   PSD of the detectors to be identical and the index $I$ can be dropped in Eq. (\ref{eq:scalar}). However, since for one case we consider different noise PSDs, we keep the subscript in this discussion to retain the generic nature.
\par 
The statistic for the $I$-th detector is, 
\begin{equation}
\Lambda_I = \max_{i, t_c} \left [(s_0^I, h^I)_I^2 + (s_{\pi/2}^I, h^I)_I^2 \right ],
\label{eq:rho}
\end{equation}
which is to be compared with the threshold.
\par
Instead of two templates $s_0^I$ and $s_{\pi/2}^I$, we find it more convenient to use a single complex template $S^I$ which combines the two together:
\begin{equation}
\tilde{S}^I (f) = \frac{1}{g_I} f^{-7/6}
\hbox{exp}\left[~ i \Psi (f;t_c,\delta_c = 0,\tau_0,\tau_3)\right].
\end{equation}
The normalization factor $g_I$ is chosen such that $(S^I,S^I)_{I}=1$. It is given by, 
\begin{equation}
g_I^2=4 \int_{f_{a}}^\infty \frac{df} {f^{7/3} S_h^I(f)} \, .
\end{equation}
In terms of $g_I$, the amplitude of the waveform $h^I$ is $\A^I = \mathcal{N} E^I g_I $. 
\par
In addition to the above normalization scheme followed for the construction of the template bank, particularly for coherent detection, where the network is treated as a whole, we also need the concept of network normalization. The total energy accessible to a network of $N$ detectors is a scalar and is given by, 
\bea
(h^I, h^I)_{\rm NW} &=& \sum_{I=1}^N (h^I, h^I)_I =  \mathcal{N}^2 \sum_{I=1}^2 g_I^2 E^*_IE^I  \no \\
                    &\equiv& |\A_{\rm NW}|^2.
\label{energy}
\eea
The quantity $\sum_{I=1}^N g_I^2 E^*_IE^I = ||E||^2$ is the ${\mathcal L}^2$ norm of $E^I$ in
${\mathcal C}^N$. The quantity $\mathcal{N}^2 g_I^2$ has the significance of the maximum possible energy accessible by the network i.e., the energy accessible when all the detectors are optimally oriented;  so $||E||^2$ denotes the ratio of the total energy received by the system to the maximum possible energy accessible. 
\par
This suggests the definition of the network-normalised signal to be
\begin{equation}
\hat{S}^I = \frac{Q^{I*}}{g_I} f^{-7/6}
\hbox{exp}\left[~ i \Psi (f;t_c,\delta_c = 0,\tau_0,\tau_3)\right].
\label{norm}
\end{equation}
where, 
\begin{equation}
Q^I = \frac{E^I}{||E||}.
\label{r2}
\end{equation}
The \emph{network vector} $\Q = (Q^1, Q^2, ..., Q^N)$ lies in ${\mathcal C}^N$ and has unit norm. This is the vector of polarization phases, which effectively `brings all the detectors to the same orientation'.
\par
For the two detector network, the data consists of two data trains, $\{ x^I(t) | I=1,2; 
\hbox{and} \; t \in [0,T] \}$ where data is taken in the time interval $[0, T]$. Assuming additive noise $n^I$ in each detector we have:
\be
x^I = h^I + n^I , ~~~ I = 1, 2. 
\ee
The noise random variables satisfy the statistical property:
\be
\langle n^I (f) n^{*}_I(f') \rangle = \h  S_h^I (f) \delta (f - f') \,, 
\label{psd}
\ee
where, the angular brackets denote ensemble average.
 In this paper we take the noise in the two detectors to be uncorrelated, i.e,
\be
\langle n^1 (f) n^{*}_2(f') \rangle = 0 \,,
\ee
which is not unjustified if the detectors happen to be geographically separated. 
For the sake of simplicity of the analysis, we  assume the noise to be both stationary and Gaussian. This is assumed in the analysis in section \ref{form} and for the simulations in section \ref{comper}. However, analysis of real data suggests it is neither - the SNRs and the $\chi^2$ are correlated \cite{TAMA}. 
\par
We conclude this section by introducing the definition of the complex correlations $C^I$ which will be particularly needed for construction of the coherent statistic. We retain the notation and definition as in paper-I which was originally introduced in \cite{PDB}. The complex conjugate of $C^I$ is denoted by $C_I^*$. It is given by:    
\begin{equation}
C_I^* = (S^I, x^I)_I = c_0^I - i c_{\pi/2}^I \,,
\end{equation}
where, $c_0^I$ and $c_{\pi/2}^I$ are the real and imaginary parts of $C^I$; they are obtained by taking the scalar products of the data $x^I$ with $s_0^I$ and $s_{\pi/2}^I$ respectively; that is,
\be
c^I_{0, \pi/2} = (s_{0, \pi/2}^I, x^I)_I \,.
\ee

\section{False alarm rate and detection probability}
\label{form}

Since in this paper we will be considering the case of two geographically separated detectors, we will consider pairs among the detectors  H1, H2, L1, Virgo and LCGT, except for the pair H1 and H2 which happen to be in the same location. In this section we will derive and present analytical formulae for the false alarm rate and detection efficiency for the coherent and coincidence strategies. In coincidence detection we consider two sub categories, one of straight forward coincidence which we call {\it naive} coincidence and a more sophisticated coincidence strategy which we call {\it enhanced} coincidence. We therefore organise our discussion as follows:

\begin{description}
\item[A.] Coherent detection 
\item[B.] Coincidence detection
\begin{description}
\item[1.] Naive coincidence detection 
\item[2.] Enhanced coincidence detection
\end{description}
\end{description}

 In the analytical formulae that we derive, we do not include all procedural steps of the numerical simulation; instead, we take a simplified route. The derived formulae are thus slightly different from those obtained from the results of the simulation. Nevertheless, the analytic formulae provide us guidelines by giving us functional forms for the false alarm and detection probabilities in terms of parameters which then can be determined from simulations. Even in such a simplified situation, 
as encountered in paper-I, where we deal with aligned detectors in the same location, the formulae depend on the unknown parameters $\Nind$ - number of statistically independent templates and $\Nwin$ -number of statistically independent templates within the error window. These quantities are difficult, if not impossible, to evaluate analytically; we therefore determine them through simulations. We find here in the case of geographically separated detectors, the quantity $\Nwin$ is significantly different from that in paper-I, because here the time-delay window becomes larger by the maximum time-delay between the two detectors. In this section, we obtain the relevant formulae in which the parameters $\Nwin$ and $\Nind$ appear implicitly. The situation will be clear from the sections which follow. 

\subsection{Coherent detection}
\label{coh}

Coherent detection involves combining data streams in a phase coherent manner so as to effectively construct a single, more sensitive detector. The maximum likelihood network statistic for two arbitrarily oriented geographically separated detectors has already been found. For the case of two arbitrarily oriented detectors the network statistic is given by \cite{PDB}: 
\begin{eqnarray}
\nonumber
\Lambda & = & ||C||^2 = |C^1|^2 + |C^2|^2 \\
& = & (c^1_0)^2 + (c^1_{\pi/2})^2 + (c^2_0)^2 + (c^2_{\pi/2})^2,
\label{twoindep}
\end{eqnarray}
where $C^I$ is the complex correlation of the $I$-th detector ($I$=1,2). \footnote{Note that in \cite{PDB} the square root of $\Lambda$ is used. But here we use $\Lambda$ because it has the $\chi^2$ statistics for Gaussian noise which has a simpler mathematical expression and thus easier to implement.} This is quite different from the coherent statistics for two coaligned detectors as it does not contain the terms involving cross-correlation between the two detectors. So, the false alarm rate and the detection efficiency changes considerably even for slightly non-aligned detectors. The comparison between the two strategies therefore must be drawn separately for non-aligned detectors, for it is not only a more general situation, it is different altogether. In \cite{mohanty} the two detector paradox regarding this abrupt change of detection statistic has been discussed and the improvement of the coherent strategy by suitably incorporating contributions from cross correlation terms has been advocated. We, however, restrict ourselves to the formalism of coherent detection as in \cite{PDB}. Any improvement to this basic formalism will enhance the performance of the coherent search and will further strengthen our results obtained in section \ref{comper}.
\par
Secondly, the correlations in the two detectors are computed at the {\it same} mass parameters $\{\tau_0, \tau_3 \}$, and at time-lags which differ at most by the light travel time $d/c$ between the detectors, where $d$ is the distance between the geographically separated detectors. This is matched filtering in which the network template is matched to the network data. The statistic is the maximum taken over the permitted time-lags and polarization phases which make up the vector $\Q$. We will see later that for enhanced coincidence, the same expression for the statistic appears, but the correlations in the two detectors are permitted to be evaluated at different points in the parameter space, constrained by a certain window or error box (which will be specified later in the text). We therefore note that this is not matched filtering.
\par
To obtain the false alarm rate and the detection efficiency, first, we concentrate on a single template. When there is no signal, $c^{1,2}_{0,\pi/2}$ are Gaussian random variables with zero mean and unit variance. So, the probability distribution of LLR can be shown to be 
\begin{equation}
p_0(\Lambda) = \frac{\Lambda}{4} \hbox{exp} \left( - \frac{\Lambda}{2} \right).
\end{equation}
The rate of false alarm for a threshold of $\Lambda^*$ is,
\bea 
P_{\rm FA}^{\rm 1 template} &=& \int_{\Lambda^*}^{\infty} d \Lambda~ p_0 (\Lambda)  \no \\
&=&  \left(1 + \frac{\Lambda^*}{2} \right) \hbox{exp} \left( - \frac{\Lambda^*}{2} \right) \,.
\eea
We argued in paper-I that when we have closely packed correlated templates, we can treat the template bank effectively as having $\Nind$ statistically independent templates. $\Nind$ is smaller than the actual number of total templates in the bank - it will turn out that $\Nind$ is substantially smaller. We will also make a simplifying assumption that the noise PSDs of the detectors are identical. With this assumption, the template placement is identical and so also is $\Nind$. In which case, for the entire template bank, the false alarm rate is,  
\begin{equation} 
P_{\rm FA} = \Nind \left(1 + \frac{\Lambda^*}{2} \right) \hbox{exp} \left( - \frac{\Lambda^*}{2} \right) \,.
\label{eq:highcoherent}
\end{equation}

On the other hand, when, there is a signal of amplitude $\A_{\rm NW}$  in the network data, the expectation value of the squared statistic is $\A_{\rm NW}^2$ and the probability distribution of the statistic is:
\begin{equation}
p_1(\Lambda) = \frac{1}{2} \left( \frac{\sqrt{\Lambda}}{\A_{\rm NW}} \right) \hbox{exp} \left[-\frac{\Lambda + \A_{\rm NW}^2}{2} \right] I_1 (A_{\rm NW} \sqrt{\Lambda}) \,,
\end{equation}
where $I_1$ is the modified Bessel function of the first kind. False dismissal occurs when in spite of the presence of the signal, the statistic $\Lambda$ falls below the threshold $\Lambda^*$ .  The false dismissal probability is given by,
\begin{eqnarray}
\nonumber
P_{\rm FD} & = &  \int_0^{\Lambda^*} \frac{d\Lambda}{2} \left( \frac{\sqrt{\Lambda}}{\A_{NW}} \right) 
\hbox{exp} \left[-\frac{\Lambda + \A_{NW}^2}{2} \right]\\ 
& \times & I_1 (A_{NW} \sqrt{\Lambda}).
\end{eqnarray}
The detection efficiency $P_{DE}$ or detection probability is then just $P_{DE} = 1 - P_{FD}$.

\subsection{Coincidence detection}
\label{coin}

In  coincidence detection, the two detectors are treated essentially in isolation. Separate lists of candidate events are prepared; a candidate event occurs in a given detector $I$ (where $I = 1, 2$), when the statistic $\Lambda_I = | C^I|^2$ computed for the detector $I$ crosses the threshold $\Lambda^*_I$ set for the detector $I$.  This procedure produces two event lists, each for one detector. In absence of prior knowledge of the signal, we may choose the same threshold $\Lambda^*$ for both detectors. The next step involves matching the lists of candidate events and obtaining pairs of events. The matching is performed by ascertaining whether the estimated parameters for a pair of candidate events, each event chosen from a separate event list, lie in a predetermined parameter window.  For a real astrophysical event, the estimated signal parameters must be consistent: ideally, the coalescence times $t_c$ must at most differ by the light travel time $d/c$, where $d$ is the distance between the detectors,  and the dynamical parameters $\tau_0, \tau_3$ must be identical. This ideal situation may only be realised in the limit of infinite SNR. However, for realistic SNRs, as are to be expected from astrophysical considerations, these constraints must be made less stringent, because the presence of noise introduces error in each parameter. The estimated parameters therefore will differ from their actual values; they must lie within an error-window $\W$. The determination of the size of this window is pivotal to this strategy. 
\par
The window $\W$ is determined as follows. Consider first the parameter $t_c$. For two geographically separated detectors, separated by distance $d$, the maximum time-delay is $d/c$. The estimated values of $t_c$ in  the two detectors can differ in addition to $d/c$ from errors due to noise. We denote the error boxes in $t_c$ due to noise by $\Dn t_c (\Lambda_1)$ and $\Dn t_c (\Lambda_2)$ in detectors 1 and 2 respectively. The error boxes depend on $\Lambda_1$ and $\Lambda_2$ for the respective events and also on the probability we can tolerate in losing an event. We choose the probability of not losing the event to be 99 $\%$ for each of the three parameters, giving a final probability of not losing an event to be $(.99)^3 \sim 0.97$ for the error window. The error box $\Dn t_c (\Lambda_1,\Lambda_2) $ is determined numerically by carrying out simulations (see next section). We assume that the errors in the two detectors are independent and so the total error $\DW t_c$ is then realised as a quadratic sum. We have:
\begin{equation}
 (\DW t_c)^2 = \left(\frac{d}{c}\right)^2 + (\Dn t_c (\Lambda_1))^2 + (\Dn t_c (\Lambda_2))^2.
\end{equation}
Similar considerations hold for the dynamical parameters $\tau_0, \tau_3$ and their corresponding error boxes: 
\begin{equation}
(\DW \tau_0)^2 = (\Dn \tau_0 (\Lambda_1))^2 + (\Dn \tau_0 (\Lambda_2))^2,
\end{equation}
and
\begin{equation}
(\DW \tau_3)^2 = (\Dn \tau_3 (\Lambda_1))^2 + (\Dn \tau_3 (\Lambda_2))^2.
\end{equation}
Let $\Delta t_c, \Delta \tau_0, \Delta \tau_3$ be the differences in the estimated parameters for the candidate pair of events corresponding to the parameters $t_c, \tau_0, \tau_3$ respectively, then we say that the parameters {\it match} if, 
\be
|\Delta t_c | \leq \DW t_c, ~~ |\Delta \tau_0 | \leq \DW \tau_0, ~~ |\Delta \tau_3 | \leq \DW \tau_3 \,.
\label{window}
\ee
Note that the dimensions of the window $\W$ are symmetric under the interchange of the detector labellings $1$ and $2$. This is important for consistency: if a trigger in detector $1$ has a window which includes a trigger in detector $2$, then the trigger in detector $2$ has a window of the same size, albeit translated, which then includes trigger $1$.  Let $\Nwin (\Lambda_1, \Lambda_2)$ be the number of templates in the window now determined by both $\Lambda_1$ and $\Lambda_2$. Note that because of the symmetry in definition of the window size, we have $\Nwin (\Lambda_1, \Lambda_2) = \Nwin (\Lambda_2, \Lambda_1)$. 
\par
Recently, a geometrical method for determining $\W$ using Fisher information matrix has been proposed which uses ellipsoidal windows instead of a rectangular ones \cite{robinson}. This scheme takes into consideration the correlation between the parameters which then reduces the false alarm rate. Incorporating ellipsoidal windows would have the effect of shifting the ROC curve for coincidence strategy a little to the left. We may estimate by how much the curve would shift. From the figure 1  shown in \cite{himan} of the ellipse in the $(\tau_0, \tau_3)$ plane describing a template one can see that the area of the ellipse is about a quarter of the corresponding rectangle. However, for the parameter $t_c$ the window must take into account the time-delay for geographically separated detectors. This is done by the translating the error ellipsoid (due to noise only)  around a trigger in one detector by the amount $\pm d/c$ and check whether the so translated ellipsoid intersects the error ellipsoid of the second detector. If $d/c$ is reasonably larger  than the errors in $t_c$ only due to noise, then there is almost no advantage to be gained in this parameter in using ellipsoidal windows. Thus the false alarm may be reduced by about a factor of 5 which is $\sim 0.7$ on the logarithmic scale (the logarithm is to base 10). Although the procedure would produce better results, for simplicity of implementation we  perform the conventional coincidence search with rectangular windows and introduce a fudge factor $\alpha \sim 1/5$ in $\Nwin$ which takes into account the effect of using ellipsoidal windows in reducing the false alarm.

\subsubsection{Naive coincidence}

The procedure for implementing naive coincidence for two arbitrarily located and oriented detectors is as follows: 

\begin{enumerate}
\item Choose the same threshold $\Lambda^*$ for the two detectors. Let $ \Lambda_I, ~ I = 1, 2$ be the individual statistics of the two detectors, prepare two candidate event lists  such that $\Lambda_I  >  \Lambda^*, ~ I = 1, 2$. 

\item Look for pairs of candidate events, each candidate event coming from a different list, such that the sets of estimated parameters match, that is, the condition (\ref{window}) is satisfied for the two events.
\end{enumerate}

If the above requirements are satisfied, announce detection. We will now use these conditions to obtain expressions for the false alarm rate and false dismissal.
\par
In paper I we have shown that for a network of two detectors, the expression for the probability of false alarm was obtained assuming a fixed window size dependent only on the threshold $\Lambda^*$. Here we have a variable window size now depending on the SNRs $\rho_I$ (or equivalently   $\Lambda_I$) of the two events and hence the derivation is more involved. However, we normally need the result at high value of the threshold $\Lambda^*$ in which case the expression for the false alarm probability assumes a simple form. 
\par
Let the statistics $\Lambda_1, \Lambda_2 > \Lambda^*$ cross the threshold $\Lambda^*$. The probability that $\Lambda_2$ lies in the infinitismal interval of size $d \Lambda_2$ for some given template is $\h e^{-\Lambda_2/2} d \Lambda_2$. Now this is the approximation we make: we assume that there is exactly one false alarm in detector 2 lying within the window; in general we could have more than one false alarm in the window, but if we assume a high value of the threshold $\Lambda^*$, then it is unlikely that one has more than one false alarm occurring within the window and thus this probability can be neglected (usually, in the literature the probability of at least one false alarm is calculated, but finally it is usually approximated to the probability of exactly one false alarm; here instead we directly compute this probability). Thus in this approximation, the probability of $\Lambda_2$ lying in an interval $[\Lambda_2, \Lambda_2 + d \Lambda_2]$ for the templates in the window is $\Nwin (\Lambda_1, \Lambda_2) \times \h e^{-\Lambda_2/2} d \Lambda_2$. This probability must be multiplied by the probability $\Nind \times \h e^{-\Lambda_1/2} d \Lambda_1$ to obtain the probability in the rectangle $[\Lambda_1, \Lambda_1 + d \Lambda_1] \times [\Lambda_2, \Lambda_2 + d \Lambda_2]$  and finally integrated from $\Lambda^*$ to $\infty$. Thus for the probability density we obtain the expression:
\be
p(\Lambda_1, \Lambda_2) = \frac{1}{4} \Nind \Nwin (\Lambda_1, \Lambda_2)   e^{-(\Lambda_1 + \Lambda_2)/2} \,.
\label{pdns}
\ee
The false alarm probability $P_{\rm FA}$ is then given by integrating this probability density over the acceptable region. Thus,
\be
P_{\rm FA} = \int_{\Lambda^*}^{\infty} d \Lambda_1 \int_{\Lambda^*}^{\infty} d \Lambda_2 ~p(\Lambda_1, \Lambda_2) \,.
\ee
For high threshold $\Lambda^*$ and also using the fact that $\Nwin (\Lambda_1, \Lambda_2)$ falls off rapidly with increasing $\Lambda_1, \Lambda_2$, it is easily seen that this expression approximates to,
\begin{equation}
P_{\rm FA} \simeq  \Nind \Nwin (\Lambda^*, \Lambda^*) \hbox{exp} \left(-\Lambda^* \right) \,.
\label{PFAnaivecoin}
\end{equation}
\par
As remarked before, $\Nwin$ depends crucially on the threshold $\Lambda^*$, because the size of the error window strongly depends upon the SNR as suggested by both Fisher information matrix considerations as well as by the simulations we perform. It also depends weakly on the location of the signal in the parameter space. However, we choose to ignore this weak dependence and treat $\Nwin$ as a function of the threshold only as in paper-I. 
Exact value of $\Nwin$ must lie between zero and the total number of templates within the parameter window. We choose $\Nwin$ to be the total number of templates within the parameter window following the practice adopted in paper-I and such a choice leads to good agreement between theoretical estimates and numerical simulations.

We now turn to the detection probability. When a signal of network amplitude $\A_{NW}$ falls on the detectors, the expected value of the statistic in each detector will be different because of differing orientations of the detectors. This information is encoded in the extended antenna pattern functions $E^I$ which depend on the detector orientations. Since the noise in the detectors is uncorrelated, the detection efficiency is just the product of the detection efficiencies of the individual detectors:
\begin{equation}
P_{\rm DE} = P_{\rm DE}^1 P_{\rm DE}^2,
\end{equation}
where, $P_{\rm DE}^I, ~I = 1, 2$ are the detection efficiencies of the individual detectors. In earlier literature, these have been calculated and here we merely state the result:
\begin{eqnarray}
\nonumber
P_{\rm DE}^I &=& \h \int_{\Lambda^*}^{\infty} d\Lambda ~\hbox{exp} \left \{ - \h  \left [ \Lambda + (\A^I )^2 \right ] \right \}\\
& \times & I_0 (\A^I \sqrt{\Lambda}) \,.
\end{eqnarray}
Note that each $\A^I, I = 1, 2$ is proportional to the extended antenna pattern functions $E^I$ respectively. The presence of the individual functions $E^I$ in the expression of detection efficiency demonstrates the weak point in this type of coincidence detection. In general, the response of the two detectors will be different because of different orientations, but since one has no \emph{a priori} knowledge of the direction of the source, one sets the same threshold for the two detectors which leads to a lot of false dismissal. In fact, for the case when the configuration is such that the response of one detector is close to zero for some particular direction to the source, actual candidate events would be missed even if the noise level is low. 
\par
The main reason behind the poor performance of the naive coincidence strategy has its roots in the sky coverage. For a particular source at a fixed distance, the intrinsic SNR of the GW emitted  depends crucially on the location and orientation of the source (the angles $\theta, \phi, \psi$ and $\iota$). For a single detector, and for a given threshold (minimum SNR), such a source would be "visible" for certain directions and orientations. When the detectors are aligned the sky coverage of the coincident detector is the same as any one of the single detectors although the coincident detector still performs better than the single detector only because of low false alarm rate. For the non-aligned case, since different parts of the sky are covered by each detector, for naive coincidence, it is the \emph{intersection} of the sky coverage of the two individual detectors which decides detection. This crucially hampers the naive coincidence strategy. Just to get a quantitative idea of the performance of this strategy, we consider a binary source  at a distance 15 Mpc. The mass of each individual star is 1.4$\msun$. The binary is taken to be optimally oriented, i.e., $\psi=\iota=0$. The location of the source in the sky is varied and the intrinsic SNR is calculated for L1, VIRGO orientations and LIGO I design sensitivity. In a simplistic scheme, if the intrinsic SNR in a given detector is above 7 for a particular direction, we take that direction of the sky as covered. We find that the coverage of a single detector is $49\%$ while in naive coincidence the detectors cover only $18\%$ of the sky. The corresponding number for the coherent strategy is $92\%$ of sky coverage.
\par
In order to counter the disadvantage of low sky coverage of this simplistic strategy, we consider a more sophisticated coincidence strategy which overcomes this problem. We describe this strategy in the next subsection.

\subsubsection{Enhanced coincidence detection}
\label{enhanced}

As we have pointed out for non-aligned detectors, the performance of coincidence strategy is poor because of low sky coverage. We adopt a more sophisticated strategy which overcomes this problem. The procedure is as follows:
\begin{enumerate}
\item Choose a low threshold $\Lambda_0^*$ and if $ \Lambda_I, ~ I = 1, 2$ are the individual statistics of the two detectors, prepare two candidate event lists  such that $\Lambda_I  >  \Lambda_0^*, ~ I = 1, 2$. 

\item Look for a pair of candidate events, the events coming from separate lists, such that the sets of estimated parameters match within the error-window; the window has already been defined in Eq. (\ref{window}).

\item  Choose the final (high) threshold $\Lambda^* > 2 \Lambda_0^*$ and construct the final statistic 
$\Lambda = \Lambda_1 + \Lambda_2$ and register detection if $\Lambda > \Lambda^*$ .
\end{enumerate}

This strategy in essence leads to increased detection efficiency. Further, this strategy also involves preparing separate candidate lists and matching candidate events, but for the case of two detectors, the detection statistic is of the same form as that of the coherent strategy. The detection statistic is, 
\begin{eqnarray}
\nonumber
\Lambda & = & ||C||^2 = |C^1|^2 + |C^2|^2 \\
& = & (c^1_0)^2 + (c^1_{\pi/2})^2 + (c^2_0)^2 + (c^2_{\pi/2})^2.
\label{twoindep2}
\end{eqnarray}
Note that although this statistic has the same form as in the coherent case, as remarked before, it is actually different: the mass parameters here do not have to be the same for the two detectors, but are only constrained to lie in the window described above. Also in $t_c$, the error box is somewhat larger than in the coherent case, where the time-lag must differ at most by $d/c$; in addition to $d/c$ there is error introduced by noise which increases the size of the error box.
\par
We first obtain the false alarm probability. We first note that the first two steps in the procedure are identical to those of naive coincidence except with the final threshold 
$\Lambda^*$ for naive coincidence replaced by the low threshold 
$\Lambda_0^*$ applied in the first step. Therefore, identical derivation follows, and hence the  probability density here is the same as before and is given by Eq. (\ref{pdns}). However, the integration region is different essentially because of the third step. Define the regions (or events) $R_1, R_2, R_{12}$ in the $(\Lambda_1, \Lambda_2)$ plane:
\bea
R_1 & = & \{(\Lambda_1, \Lambda_2) | \Lambda_1 > \Lambda_0^*,~~ 0 \leq \Lambda_2 < \infty \} \,, \no \\
R_2 & = & \{(\Lambda_1, \Lambda_2) | \Lambda_2 > \Lambda_0^*,~~ 0 \leq \Lambda_1 < \infty \} \,, \no \\ 
R_{12} &=& \{(\Lambda_1, \Lambda_2) | \Lambda_1 + \Lambda_2 > \Lambda^* \} \,. \no \\
\eea 
Let the region $R = R_1 \cap R_2 \cap R_{12}$ be the intersection of all the three regions. Note that since $\Lambda^*$ is chosen greater than $2 \Lambda_0^*$ we have $R$ as a proper subset of $R_1 \cap R_2$. The region `cut out' from $R_1 \cap R_2$ is an isosceles right angled triangle, we call $\Delta$. Thus $R_1 \cap R_2 = R + \Delta$ where $+$ denotes disjoint union of the sets. The false alarm probability is given by integrating the probability density given by Eq. (\ref{pdns}) over the region $R$. Thus:
\be
P_{\rm FA} = \int_{R} d \Lambda_1 d \Lambda_2~ p (\Lambda_1, \Lambda_2) \,.
\ee  
We may be able to approximate this expression by writing $R = R_1 \cap R_2 - \Delta$ which leads to:
\be
P_{\rm FA} = P (R_1 \cap R_2) - P (\Delta) \,.
\ee
If again $\Nwin$ is a rapidly decreasing function of both its arguments, we can again write:
\be
P (R_1 \cap R_2) \approx \Nind \Nwin (\Lambda_0^*, \Lambda_0^*) e^{- \Lambda_0^*} \,.
\ee
On the other hand, $P(\Delta)$ becomes
\begin{widetext}
\bea
P(\Delta)&=&\int_\Delta dx_1 dx_2~ p (x_1, x_2) \nonumber\\
&=&\frac{N_{\rm ind}}{4}
\int_{\Lambda_0^*}^{\Lambda^*-\Lambda_0^*}dx_1~ e^{-\frac{1}{2}x_1}
\int_{\Lambda_0^*}^{\Lambda^*- x_1}dx_2 ~\Nwin(x_1,x_2)~e^{-\frac{1}{2}x_2}
\nonumber\\
&\simeq& \frac{N_{\rm ind}}{4}~\Nwin(\Lambda_0^*,\Lambda_0^*)
\int_{\Lambda_0^*}^{\Lambda^*-\Lambda_0^*}dx_1 ~2e^{-\frac{1}{2}x_1}~
\left(e^{-\frac{1}{2}\Lambda_0^*}-e^{-\frac{\Lambda^*-x_1}{2}}\right)
\nonumber\\
&=&N_{\rm ind}~\Nwin(\Lambda_0^*,\Lambda_0^*)~\left[e^{-\Lambda_0^*}-e^{-\frac{\Lambda^*}{2}}
\left(\frac{\Lambda^*}{2}-\Lambda_0^*+1\right)
\right].
\eea
Thus, we have
\bea
P_{\rm FA}\simeq N_{\rm ind}~\Nwin(\Lambda_0^*,\Lambda_0^*)
~e^{-\frac{\Lambda^*}{2}}~\left(\frac{\Lambda^*}{2}-\Lambda_0^*+1\right).
\label{eq:FAenh}
\eea
\end{widetext}

We now turn to the detection probability. We assume a signal in the detectors with amplitudes $\A^I, ~I = 1, 2$ and true signal parameters $\lambda^\mu_0$, where 
$\lambda^\mu = \{t_c, \tau_0, \tau_3 \}$. The true signal parameters $\{\tau_0, \tau_3 \}$ will be the same for the two detectors while the parameter $t_c$ will differ by the light travel time depending on the direction of the wave. Thus $\lambda^\mu_0$ must be interpreted according to the context. However, because of noise in the detectors, the detector statistics $\Lambda_1$ and $\Lambda_2$ in general will be evaluated at points in the parameter space different from $\lambda^\mu_0$, say at, $\lambda^\mu_1$ and $\lambda^\mu_2$, which will lie  close to the true signal parameters $\lambda^\mu_0$, if the signal amplitude is sufficiently high. We now compute the probability density function (pdf) for the statistic $\Lambda_1$ in the limit of high amplitude for detector 1. Similar arguments will hold for the pdf for detector 2. In the limit of high amplitude we may assume the errors in the estimated parameters in detector 1 to be Gaussian distributed with a covariance matrix $C_1^{\mu \nu}$. Also the  amplitude parameter of the signal will be be reduced by the ambiguity function $\H (\Delta \lambda^\mu_1)$, where we assume that the ambiguity function only depends on the difference in signal parameters $\Delta \lambda^\mu_1 = \lambda^\mu_1 - \lambda^\mu_0$. The ambiguity function is normalised to $\H (0) = 1$ which is also its maximum value. (Since we have assumed identical noise PSD for the two detectors, the ambiguity functions for the two detectors are the same and accordingly we have omitted the detector  index $I$ for the ambiguity function.) The joint pdf for the statistic $\Lambda_1$ and signal parameters $\lambda^\mu_1$ is given by,
\be
p_1 (\lambda^\mu_1; \Lambda_1) = f(\lambda^\mu_1)~ g (\lambda^\mu_1; \Lambda_1) \,,
\ee
where,
\be
f(\lambda^\mu_1) = \frac{1}{(2 \pi)^{3/2}(\det C_1)^\h}
~e^{- \h [C_1^{-1}]_{\mu \nu} \Delta \lambda^\mu_1 \Delta \lambda^\nu_1} \,, 
\ee 
and,
\bea
g (\lambda^\mu_1; \Lambda_1) &=& \h e^{- \h [\Lambda_1 + (\A^1)^2 \H^2 (\Delta \lambda^\mu_1)]} \no \\
& \times &I_0 [\A^1 ~\H (\Delta \lambda^\mu_1)~ \sqrt{\Lambda_1}].
\eea
A similar expression obtains for detector 2 with the index 1 replaced by the index 2. 
\par
We now note that the covariance matrix scales as the inverse of the square of the amplitude or $C^{\mu \nu}_1 \sim C^{\mu \nu} (\A^1 = 1) / (\A^1)^2$ and in the limit of $\A^1 \longrightarrow \infty$, the function $f(\lambda^\mu_1)$ tends to a delta function centered at the signal parameters $\lambda^\mu_0$. Now we integrate over the signal parameters 
$\lambda^\mu_1$ or marginalise over them. Assuming high amplitude $\A^1$, we expand $g$ up to the second order in $\Delta \lambda^\mu_1$ around $\lambda^\mu_0$, and the result after integration is:
\be
p_1 (\Lambda_1) = p_1 (\Delta \lambda^\mu = 0; \Lambda_1) + \frac{k^1_{\mu \nu} C^{\mu \nu}_1 (\A^1 = 1)}{(\A^1)^2} \,,
\ee
where $k^1_{\mu \nu}$ are constants, namely, the second derivatives of $g$ with respect to 
$\Delta \lambda^\mu$ evaluated at $\Delta \lambda^\mu = 0$. This means that the pdf for detector 1 is essentially given by the first term plus corrections of order $o (1 /(\A^1)^2)$. These corrections are small if $\A^1 >> 1$. Or writing this result explicitly:
\be
p_1 (\Lambda_1) = \h e^{- \h [\Lambda_1 + (\A^1)^2]} I_0 (\A^1 \sqrt{\Lambda_1}) + o (1 /(\A^1)^2) \,.
\ee
For the detector 2 we obtain an identical result with $1$ replaced by $2$ on the RHS. Finally, the pdf for the two detectors is obtained by simply multiplying the pdfs of the individual detectors since the statistics $\Lambda_1$ and $\Lambda_2$ are independent. Thus the detection probability or efficiency is obtained by integrating the product of the pdfs over the region $R$ and is given by,
\be
P_{DE} = \int_{R} p_1 (\Lambda_1)~ p_1 (\Lambda_2)~ d \Lambda_1 d \Lambda_2 \,.
\ee
\par
In this strategy the sky coverage is better than the naive coincidence because $\Lambda_0^*$ is chosen much smaller than the threshold $\Lambda^*$ chosen in naive coincidence strategy. Thus we expect the performance to be better than the naive coincidence. In the next section we compare the performances of all the strategies.  

\section{Comparing performances}
\label{comper}

In the previous section we have derived analytical formulae for plotting the ROC curves. 
However, the quantities, $\Nind$ and $\Nwin$, are unknown and we need to determine them by numerical simulation before we can actually plot the curves. On the other hand, as we shall see below, the false alarm probability and the detection efficiency from the numerical simulation do not necessary agree with theoretical formulae perfectly due to various practical reasons. Thus, we have to resort to numerical simulations for evaluating quantitatively the performances of each detection strategy. 
Nevertheless, the analytical formulae provide guidance towards deriving useful fitting formulae from the simulation which are then used to plot the ROC curves. In paper-I we have already discussed this matter in detail. However, we include a description for making the paper self contained.
\vspace{.2in}

\subsection{Simulation method}

In this paper, we take the noise PSDs of all the detectors to be identical. We use the design sensitivity of initial LIGO \cite{DIS} for all the detectors.
\begin{widetext}
\begin{eqnarray}
\nonumber
S_h(f) & = & S_0 \left[ (4.49x)^{-56} +0.16x^{-4.52} + 0.52 + 0.32x^2 \right] \qquad  f \geq f_s \\
& = & \infty \qquad \hbox{otherwise},
\end{eqnarray}
\end{widetext}
where, $x = f/f_k$, $f_k = 150$ Hz, $f_s = 40$ Hz,  and $S_0 = 9 \times 10^{-46}$ Hz$^{-1}$. 
For computing the SNR (scalar product) we set $f_l=f_s$. We take the upper cut-off frequency $f_u$ to be 1024 Hz and the data are sampled at 2048 Hz in accordance with the Nyquist theorem. 
This choice of the noise PSD is simply to save the computation time of the simulation.
If we take the PSD of advanced LIGO, the lower cut off frequency $f_s$ becomes around $10$ Hz.
The length of the signal in the bandwidth becomes longer which makes the computation time of the matched filtering longer. Even if we were to adopt the PSD of advanced LIGO, we do not expect much  qualitative change in the results. We consider 4 detector pairs listed in Table \ref{tab:pairs}. 

\begin{table}
\begin{center}
\begin{tabular}{cc}
\hline 
Pair & Light travel time [ms] \\
\hline 
L1-V1 & 26.42 \\
L1-H1 & 10.02 \\
L1-K1 & 32.47 \\
V1-K1 & 29.18 \\
\hline
\end{tabular}
\caption{The light travel time between pairs of detectors.}
\label{tab:pairs}
\end{center}
\end{table}

Using 3PN restricted waveform, we place the templates 25 ms apart in the $\tau_0$ direction 
and 19 ms apart in the $\tau_3$ direction. This corresponds to a maximum mismatch of $3\%$. 
We have taken a rectangle in the parameter space with 25 templates along each of the $\tau_0, \tau_3$ directions, totally giving us 625 templates in the $\tau_0, \tau_3$ parameters. 
We take a data train 32 seconds long and sample it at  2048 Hz. We generate 32 seconds Gaussian data and repeat the coherent and coincident searches for $N_{\rm sim}$ number of times. We test for various values of $N_{\rm sim}$, and we find that $N_{\rm sim}=20000$ is sufficient to give reliable estimates of the false alarm rate. Therefore the total data length of the simulation is $6.4\times 10^5$ seconds.

In the coherent searches, we first compute the statistics from each detector,
$\Lambda_1(t_c,\tau_0,\tau_3)\equiv(c_0^1)^2+(c_{\pi/2}^1)^2$ and 
$\Lambda_2(t_c,\tau_0,\tau_3)\equiv(c_0^2)^2+(c_{\pi/2}^2)^2$ (i.e., Eq.(\ref{twoindep})), which are functions of $t_c, \tau_0$ and $\tau_3$. We compute $\Lambda_1+\Lambda_2$ 
and take the maximum over the possible time delay and over the mass template space.
We obtain the following statistic,
\begin{eqnarray}
&& \Lambda(t_c) =   \nonumber \\
&&\max_{\tau_0,\tau_3}\max_{|t_d|\leq d/c}
\left[\Lambda_1(t_c,\tau_0,\tau_3)+\Lambda_2(t_c + t_d,\tau_0,\tau_3)\right],
\nonumber\\
\label{cohmax1}
\end{eqnarray}
where $d$ is the distance between the detectors and $c$ the speed of light.
Since the coalescence time $t_c$ is sampled at 2048 Hz, the samples $\Lambda (t_c)$ (sampled at the same rate as $t_c$) are correlated. In order to remove the correlation, following the procedure in Paper I, we divide the data train into $\delta t=15.6$ms short data trains, where the $i-$th short train starts at time $t_{(i)}=15.6 {\rm ms} \times i$. In each short data train, we take the maximum of $\Lambda(t_c)$,
\begin{equation}
\Lambda_{(i)}=\max_{t_{(i)}\leq t_c\leq t_{(i+1)}} \Lambda(t_c). \label{cohmax2}
\end{equation}
$\{\Lambda_{(i)}\}$ defines the trigger list of the simulation.

In the coincident analysis, we first compute the statistics of each detector and 
take the maximum over the mass template space. As in the coherent analysis, we divide the data train 
into $\delta t=15.6$ms short data trains, and take the maximum of the statistics in each data train.
We obtain the triggers from each detector defined as 
\begin{eqnarray}
\Lambda_{I(i)}&=&\max_{t_{(i)}\leq t_c\leq t_{(i+1)}} \Lambda_{I}(t_c), \label{coinmax1}\\
\Lambda_{I}(t_c)&\equiv&\max_{\tau_0,\tau_3}\Lambda_I (t_c,\tau_0,\tau_3), \quad (I=1,2) \,. \label{coinmax2}
\end{eqnarray}
In the naive coincidence detection, for each $\Lambda_{1(i)} > \Lambda_0^*$, we select a maximum of 
$\Lambda_{2(j)}$  which satisfies the coincident conditions, namely, Eq.(\ref{window}) and secondly 
$\Lambda_{2(j)} > \Lambda_0^*$. If such a $\Lambda_{2(j)}$ exists, the pair $\{\Lambda_{1(i)},\Lambda_{2(j)}\}$ define the trigger list of the naive coincidence detection. From this trigger list, we compute a enhanced coincident statistic defined as
\begin{eqnarray}
\Lambda&=&\Lambda_{1(i)}+\Lambda_{2(j)}.
\end{eqnarray}

\subsection{Coincidence windows}

To determine the size of windows for coincident analysis, one way is to use the Fisher information matrix for this purpose. However, it has been shown that at low SNR,  SNR $\lsim 10$, the Fisher information matrix grossly underestimates the size of the window \cite{BSD}. 
So, we determine empirically the number of templates falling within the error-window $\W$ 
such that the signal is detected $99\%$ of the time. 
Since we have three parameters, the coalescence time $t_c$ and the dynamical parameters $\tau_0$ and $\tau_3$, on the average, the fraction of signals detected will be $(.99)^3 \sim 0.97$. 
The simulations are carried out for a single detector for SNR($\equiv \sqrt{\Lambda_I}$) ranging from 4 to 25. 
Determination of $\W$ at low SNR is difficult because the signal tends to get 
overwhelmed by false alarm triggers - in most cases, instead of the signal we pick up false alarms.  
Since the false alarm triggers occur anywhere in the parameter space at random, it becomes impossible to decide on the window. So, we restrict ourselves only to those signals for which $SNR \geq 4$. For SNR$> 25$, we take the same window size as SNR$=25$, since the parameter estimation
accuracy becomes nearly constant at large SNR due to the finite mesh size of the mass template space
and the finite sampling rate. This constant window size at SNR$>25$ does not affect our analysis very much because we will use the false alarm probability and the detection probability 
corresponding $\Lambda(={\rm SNR}^2)\lesssim 200$. The parameter estimation errors from the simulation are summarized in Table \ref{win}. It is useful to deduce a fitting formula of Table \ref{win}. We observe that the estimated error scales roughly inversely as the SNR which we denote by $\rho$. Thus, we may write: 
\begin{eqnarray}
\Delta_n t_c&=&\frac{a_{t_c}}{\sqrt{\Lambda}}, \\
\Delta_n \tau_0&=&\frac{a_{\tau_0}}{\sqrt{\Lambda}}, \\
\Delta_n \tau_3&=&\frac{a_{\tau_3}}{\sqrt{\Lambda}},
\end{eqnarray}
for $\rho\leq 25$ with $a_{t_c}=98.7 {\rm ms}, \quad a_{\tau_0}=1895.4{\rm ms}$, and $\quad a_{\tau_3}=1386.0{\rm ms}$. The window size is then given assuming the box window as: 
\begin{eqnarray}
&&(\Delta_{\cal W}t_c)(\Delta_{\cal W}\tau_0)(\Delta_{\cal W}\tau_3) \nonumber\\
&&~~~=\left[\frac{d^2}{c^2} + \frac{a_{t_c}^2}{\Lambda_1}+\frac{a_{t_c}^2}{\Lambda_2}\right]^{1/2} \nonumber\\
&&~~~\times \left[\frac{a_{\tau_0}^2}{\Lambda_1}+\frac{a_{\tau_0}^2}{\Lambda_2}\right]^{1/2}
\left[\frac{a_{\tau_3}^2}{\Lambda_1}+\frac{a_{\tau_3}^2}{\Lambda_2}\right]^{1/2},
\end{eqnarray}
where $d$ is the distance between the two detectors. As discussed in paper-I, the decorrelation length of $t_c$, $\tau_0$ and $\tau_3$ can be taken as $\sim 15.8$ms, $220$ms and $80$ms. 
The number of independent templates in the coincidence window size is thus estimated roughly as
\begin{eqnarray}
N_{\rm win}&\sim& N_{\rm win}^{(0)}
\left[\frac{1}{a_{t_c}^2}\left(\frac{d}{c}\right)^2+\frac{1}{\Lambda_1}+\frac{1}{\Lambda_2}\right]^{1/2} \nonumber\\
&& \times \left[\frac{1}{\Lambda_1}+\frac{1}{\Lambda_2}\right].
\label{eq:Nwinfit}
\end{eqnarray}
\be
N_{\rm win}^{(0)} \equiv \frac{a_{t_c}}{15.8{\rm ms}}\frac{a_{\tau_0}}{220{\rm ms}}\frac{a_{\tau_3}}{80{\rm ms}}
\sim 932.4.
\ee
Note however that we do not use the value of $N_{\rm win}^{(0)}$ explicitly. We use only the functional form of Eq.(\ref{eq:Nwinfit}) to obtain the formulae for the false alarm rate.

\begin{table}
\begin{tabular}{|c|c|c|c|}
\hline
SNR & $\Dn t_c$ & $\Dn \tau_0$ & $\Dn \tau_3$ \\
\hline
4 & 24.2 & 476 & 314 \\
5 & 23.1 & 413 & 289 \\
6 & 19.6 & 351 & 260 \\
7 & 16.0 & 300 & 234 \\
8 & 11.6 & 239 & 188 \\
9 & 9.8  & 194 & 157 \\
10 & 8.0 & 162 & 129  \\
11 & 6.3 & 136 & 108 \\
12 & 5.9 & 126 & 101 \\
13 & 5.3 & 110 & 89 \\
14 & 4.9 & 109 & 87 \\
15 & 4.8 &  98 & 80 \\
20 & 4.1 &  82 & 70 \\
$>$25 & 3.6 &  79 & 66 \\
\hline
\end{tabular}
\caption{Average errors in parameters $t_c$, $\tau_0$ and $\tau_3$ in units of milliseconds due to detector noise for SNR$\equiv \sqrt{\Lambda_I}$ ranging from 4 to 25. The size of the error box corresponds to $99\%$ detection. For SNR$> 25$, we take the same window size as SNR$=25$.}
\label{win}
\end{table}

\begin{figure}[htbp]
\begin{center}
\mbox{(a) L1-V1}
\includegraphics[width=.5\textwidth]{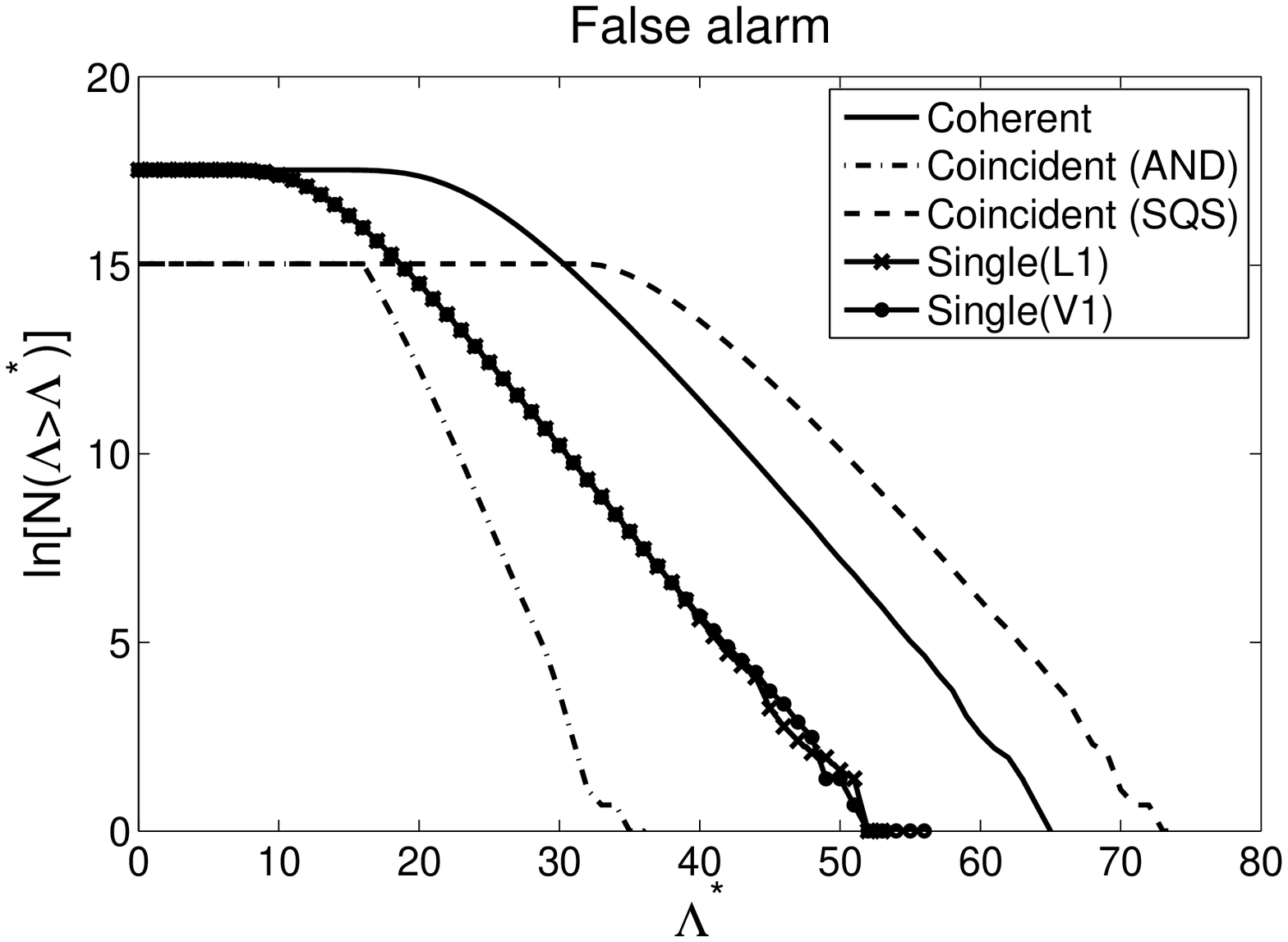}
\mbox{(b) L1-H1}
\includegraphics[width=.5\textwidth]{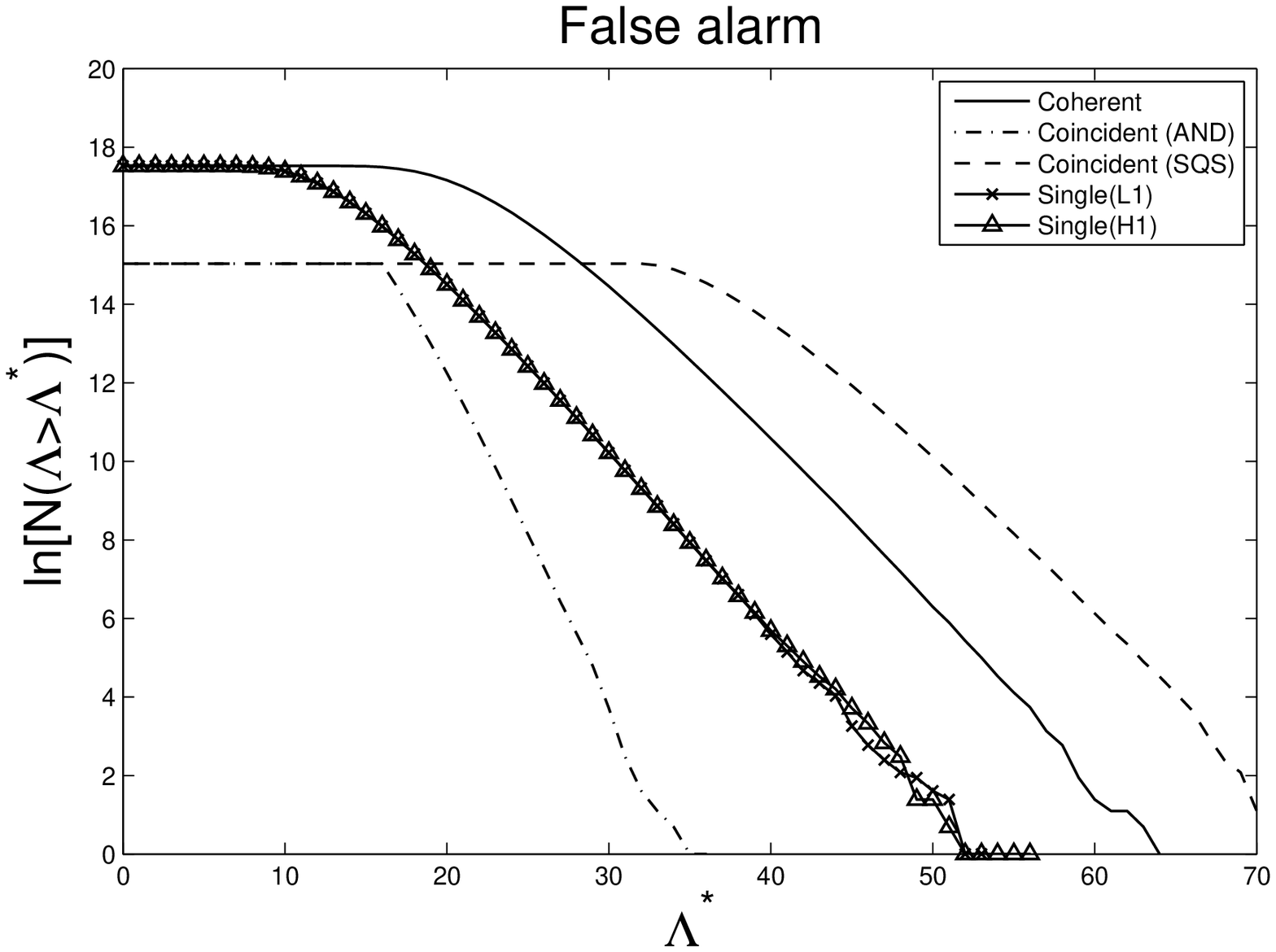}
\mbox{(c) L1-K1}
\includegraphics[width=.5\textwidth]{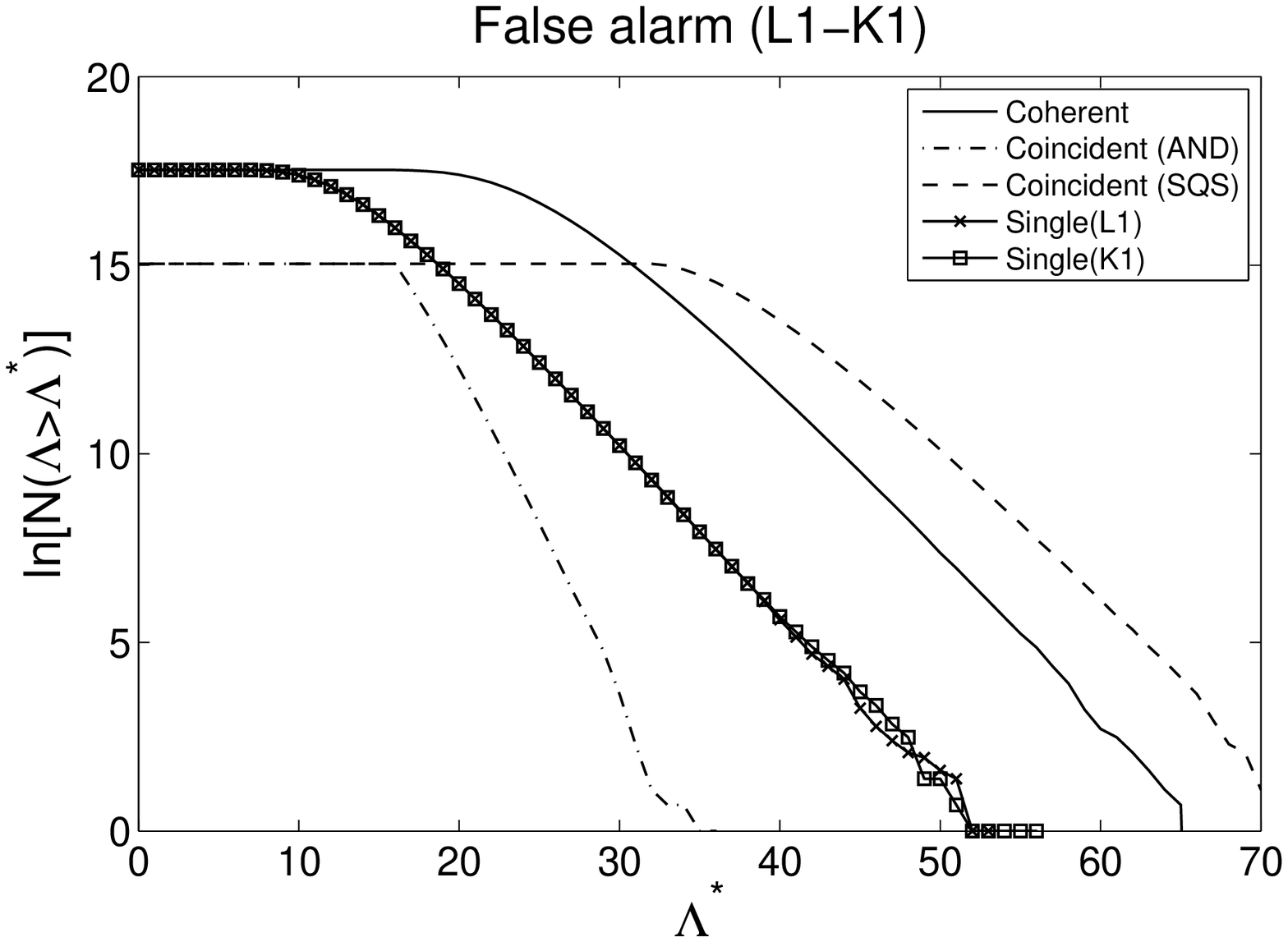}
\mbox{(d) V1-K1}
\includegraphics[width=.5\textwidth]{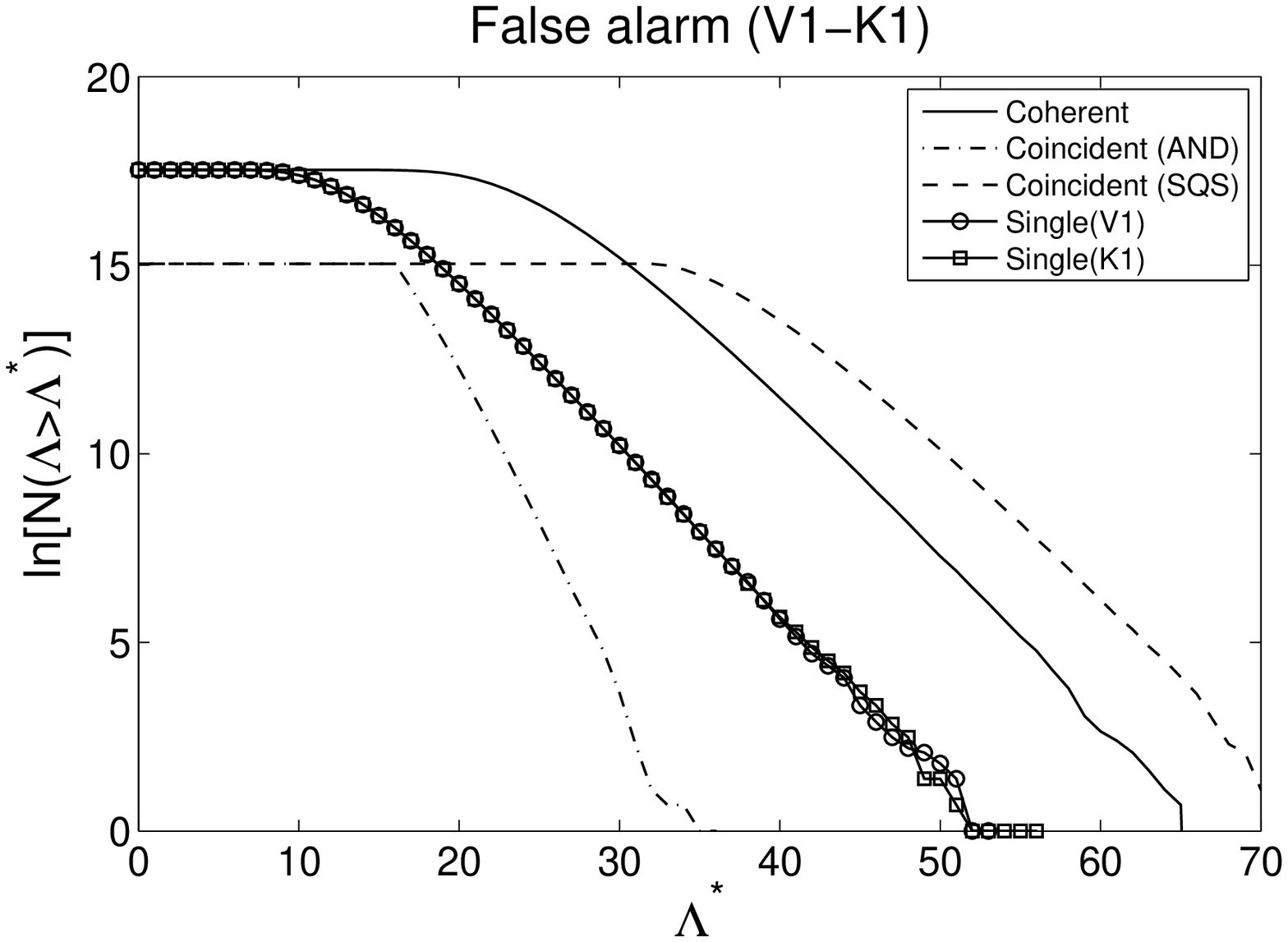}
\caption{The number of false alarm triggers as a function of the threshold for various 
detector combinations.}
\label{fig:falseall} 
\end{center}
\end{figure}

\subsection{Estimating the false alarm rate}

The number of false alarm triggers as a function of the threshold from simulations is shown 
in Fig.\ref{fig:falseall}.
In these figures, and in the rest of the figures in this paper, "coincident (AND)" and "coincident (SQS)"
indicate the naive coincident strategy and the enhanced coincident strategy respectively.
We extrapolate these plots to larger threshold in order to obtain the false alarm probability
corresponding to a much larger data set. In this extrapolation, the analytic formulae in the previous section are helpful. In the single detector case, the number of false alarms, $N_{\rm FA}(\Lambda>\Lambda^*)$ is given analytically as,
\begin{eqnarray}
N_{\rm FA}(\Lambda>\Lambda^*)=N_{\rm ind} e^{-\frac{1}{2}\Lambda^*}.
\end{eqnarray}
The factor in the exponential, namely, $-1/2$ does not necessary hold in a simulation. We thus replace $-1/2$ and also $\ln N_{\rm ind}$ with parameters $a$ and $b$ to be determined from the simulations. We take the logarithm of the above equation and write:
\begin{equation}
\ln N_{\rm FA}(\Lambda>\Lambda^*) = -a \Lambda^* + b \,,
\end{equation}
where $a$ is the slope and $b$ is the intercept if we plot $\ln N_{\rm FA}(\Lambda>\Lambda^*)$ versus $\Lambda^*$. From simulations we plot the required curve and read off $a$ and $b$.

We apply similar methods to coherent and coincident cases. In the analytical formula
for the coherent case, the number of false alarms falls off as Eq. (\ref{eq:highcoherent}):
\begin{equation}
\ln N_{\rm FA}(\Lambda>\Lambda^*) = -\frac{1}{2} \Lambda^* + \ln \left(1+\frac{\Lambda^*}{2}\right) + \ln \Nind,
\label{coh}
\end{equation}
In the case of naive coincidence, the number of false alarms is given by (Eq.(\ref{PFAnaivecoin})).
We use Eq.(\ref{eq:Nwinfit}) for $\Nwin$ and obtain, 
\begin{eqnarray}
\ln N_{\rm FA}(\Lambda>\Lambda^*) 
&=& - \Lambda^* + \frac{1}{2}\ln\left[\frac{1}{a_{t_c}^2} \left(\frac{d}{c}\right)^2+\frac{2}{\Lambda^*} \right] 
\nonumber\\
&&+ \ln\left(\frac{2}{\Lambda^*}\right)  
+ \ln (\Nind N_{\rm win}^{(0)}). \nonumber\\
\,
\label{coin1}
\end{eqnarray}
In the case of enhanced coincidence, we have from Eq.(\ref{eq:FAenh}),
\begin{eqnarray}
\ln N_{\rm FA}(\Lambda>\Lambda^*) &=& 
- \frac{1}{2} \Lambda^* + \ln \left(\frac{\Lambda^*}{2}-\Lambda^*_0+1\right)
\nonumber\\
&&+\ln \left(\Nind N_{\rm win}(\Lambda^*_0,\Lambda^*_0)\right),
\nonumber\\
\,\label{coin2}
\end{eqnarray}
where $\Lambda^*_0=4^2$. The dominant term which determines the slope of each curve is $-\Lambda^*/2$ 
in Eq.(\ref{coh}), $-\Lambda^*$ in Eq.(\ref{coin1}), and $-\Lambda^*/2$ in Eq.(\ref{coin2}).
We replace these factors with $a$, and unknown terms $\ln \Nind$ in Eq.(\ref{coh}),
$\ln (\Nind N_{\rm win}^{(0)})$ in Eq.(\ref{coin1}) and 
$\ln (\Nind N_{\rm win}(\Lambda_0,\Lambda_0))$ with $b$ respectively.
We then determine $a$ and $b$ from the simulations.
By fitting the curves obtained from simulations we determine $a$ and $b$ in the four cases: 
(i) the single detector case, (ii) the coherent case, (iii)  the naive coincidence case and finally 
(iv) the enhanced coincidence case. 
The results are listed in Table \ref{tab:fitFA} in this order for the four detector pairs 
(L1, V1), (L1, H1), (L1, K1) and (V1, K1). 
We can see that in all cases, the slope $a$ is slightly smaller than  theoretical value. 
This is mainly because of the maximization over the coalescence time and the two mass parameters  described in Eq.(\ref{cohmax1})-(\ref{coinmax2}). 
When we take the maximum of $\Lambda$, we tend to pick up large $\Lambda$ events.
This produces the deviation from the simple theoretical curve. 
We have confirmed this by performing simulations in which no maximization is done. 
When no maximization is done, we obtain the slope $a$ which agrees with the theoretical value
in all of cases from (i) to (iv). 
In this paper, the maximization about $t_c$, the time delay, and two masses are essential 
to eliminate the statistical correlation between nearby templates. 
Thus, the deviation of the slope $a$ from the theoretical value is not unnatural. 
Further, in the analysis with real data, this type of maximization process is usually carried out. 
Thus, such a deviation is more realistic than the values obtained from simple theoretical analysis.

\begin{table}
\centering
\begin{tabular}{|c|c|c|c|}
\hline
L1-V1 & Fitting region & $a$ & $b$\\
\hline
Single &   30 $\leq \Lambda^* \leq$ 50 & -0.44567 & 23.5293 \\
Coherent  &   40 $\leq \Lambda^* \leq$ 60 & -0.4551 & 26.6683 \\
Naive coincidence & 23 $\leq \Lambda^* \leq$ 33 & -0.89681 & 34.0749 \\
Enhanced coincidence &  50 $\leq \Lambda^* \leq$ 67 & -0.46412 & 31.2254 \\
\hline
\end{tabular}
\hbox{(a) L1-V1}
\begin{tabular}{|c|c|c|c|}
\hline
L1-H1 & Fitting region & $a$ & $b$\\
\hline
Single &   30 $\leq \Lambda^* \leq$ 50 & -0.44624 & 23.5448 \\
Coherent  &   40 $\leq \Lambda^* \leq$ 60 & -0.4668 & 26.3345 \\
Naive coincidence & 23 $\leq \Lambda^* \leq$ 33 & -0.84920 & 33.1193 \\
Enhanced coincidence &  50 $\leq \Lambda^* \leq$ 67 & -0.46104 & 31.0549 \\
\hline
\end{tabular}
\hbox{(b) L1-H1}
\begin{tabular}{|c|c|c|c|}
\hline
L1-K1 & Fitting region & $a$ & $b$\\
\hline
Single &   30 $\leq \Lambda^* \leq$ 50 & -0.44596 & 23.5328 \\
Coherent  &   40 $\leq \Lambda^* \leq$ 60 & -0.45420 & 26.8005 \\
Naive coincidence & 23 $\leq \Lambda^* \leq$ 33 & -0.90159 & 34.0837 \\
Enhanced coincidence &  50 $\leq \Lambda^* \leq$ 67 & -0.46407 & 31.2229 \\
\hline
\end{tabular}
\hbox{(c) L1-K1}
\begin{tabular}{|c|c|c|c|}
\hline
V1-K1 & Fitting region & $a$ & $b$\\
\hline
Single &   30 $\leq \Lambda^* \leq$ 50 & -0.43849 & 23.2738 \\
Coherent  &   40 $\leq \Lambda^* \leq$ 60 & -0.45480 & 26.7335 \\
Naive coincidence & 23 $\leq \Lambda^* \leq$ 33 & -0.90073 & 34.1248 \\
Enhanced coincidence &  50 $\leq \Lambda^* \leq$ 67 & -0.46410 & 31.2301 \\
\hline
\end{tabular}
\hbox{(d) V1-K1}
\caption{The results of the fitting of the number of false alarm triggers in
various detector pairs.}
\label{tab:fitFA}
\end{table}

\subsection{The ROC curves}

\begin{figure}[htbp]
\centering
\mbox{(a) L1-V1}
\includegraphics[width=.5\textwidth]{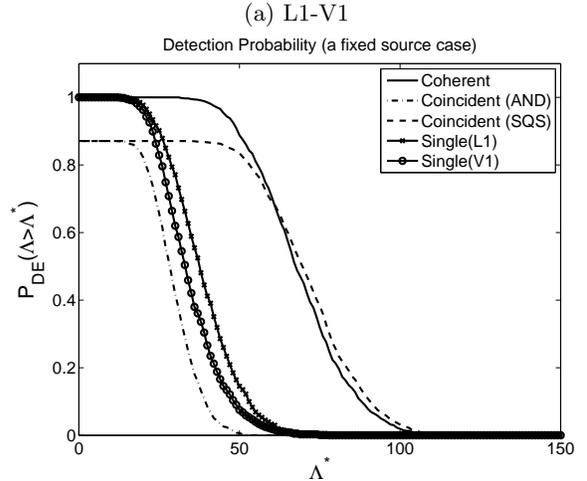} 
\caption{The detection probability for coherent, coincidence and single detector for a fixed source in the LIGO (L1) and VIRGO (V1) case. The source is a 1.4-1.4 $M_\odot$ binary located at 35 Mpc with $\psi=0$, $\iota=0$, $\theta=1.0$ rad and $\phi=0.8$ rad. The amplitude of the signal at each detector is $\Lambda=32.2$(L1) and $\Lambda=28.3$(V1) respectively.}
\label{fig:DEfitL1V1}
\end{figure}

After ascertaining the formulae for the false alarm, we proceed to plot the ROC curves. 
We first consider a hypothetical source such that the mass of each star in the binary is equal to 1.4$\msun$. We consider the detector pair (L1, V1). We take the source to be located at 35 Mpc with the inclination angle and the polarization angle as $0$. The direction to the source is taken to be described by the polar angles $\theta=1.0$ rad and $\phi=0.8$ rad in the Earth centered 
coordinate system. The amplitude of the signal in the noise free situation is then given by ${\cal A}^2=32.2$ for L1 and ${\cal A}^2=28.3$ for V1. The detection probability for this case is shown in 
Fig.\ref{fig:DEfitL1V1}. Note that the detection probabilities in Fig.\ref{fig:DEfitL1V1} are slightly worse than the theoretical values obtained in Section III. This is because, in the simulation, we use a  discrete time step and also discrete set of mass parameters values which then produce a mismatch between the signal and templates. Thus, the amplitude of the signal detected is smaller than injected value.

\begin{figure}[htbp]
\centering
\includegraphics[width=.5\textwidth]{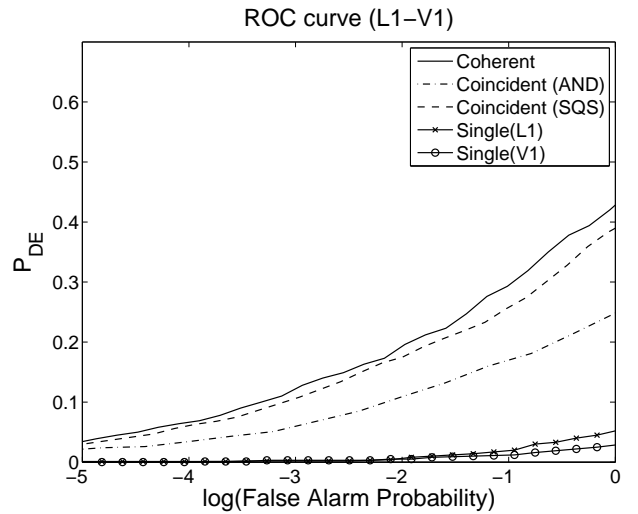} 
\caption{The ROC curves for coherent, coincidence and single detector for a fixed source
in the LIGO (L1) and VIRGO (V1) case. The source is a 1.4-1.4 $M_\odot$ binary located at 35 Mpc with 
$\psi=0$, $\iota=0$, $\theta=1.0$ rad and $\phi=0.8$ rad. The amplitude of the signal at each detector is
$\Lambda=32.2$(L1) and $\Lambda=28.3$(V1) respectively.}
\label{fig:ROCfixL1V1}
\end{figure}

From the detection probability and the false alarm rate obtained in the previous subsection, 
we plot the ROC curves. We use the fitting formulae for the false alarm rate. We assume an year's worth of observation period and templates in each individual mass ranging from $1M_\odot$ to $40M_\odot$. The total number of time samples is then $6.1\times 10^{10}$ and the number of mass templates is $1.2\times 10^4$. On the other hand, the number of mass templates in the simulation is 625 and the total number of time samples is $32\times2048\times2\times10^4$ $=1.3\times 10^9$. Thus, the false alarm probability for one year observation period with template mass range of $1-40M_\odot$ is found by scaling the false alarm probability by the factor: 
$$
\frac{1.2\times 10^4}{625}\frac{6.1\times 10^{10}}{1.3\times 10^9}=900.9.
$$
We therefore add $\ln(900.9)=6.8$ to $b$ in Table \ref{tab:fitFA} to the relevant L1-V1 part of the Table. As discussed in Section III.B, the coincidence window for two coincident strategies can be improved using ellipsoidal windows with a factor of 5 reduction in the false alarm rate. 
We thus subtract $\ln(5)$ from $b$ for naive and enhanced coincidence strategies in Table 
\ref{tab:fitFA}. We show the ROC curves in Fig. \ref{fig:ROCfixL1V1}. We find that the coherent case
gives the largest detection probability. 
It is about 70 to 80\% larger than the naive coincident case. 
The relative difference between the coherent case and the enhanced coincident case 
is about 10 to 20\%. 

\begin{figure}[htbp]
\centering
\mbox{(a) L1-V1}
\includegraphics[width=.5\textwidth]{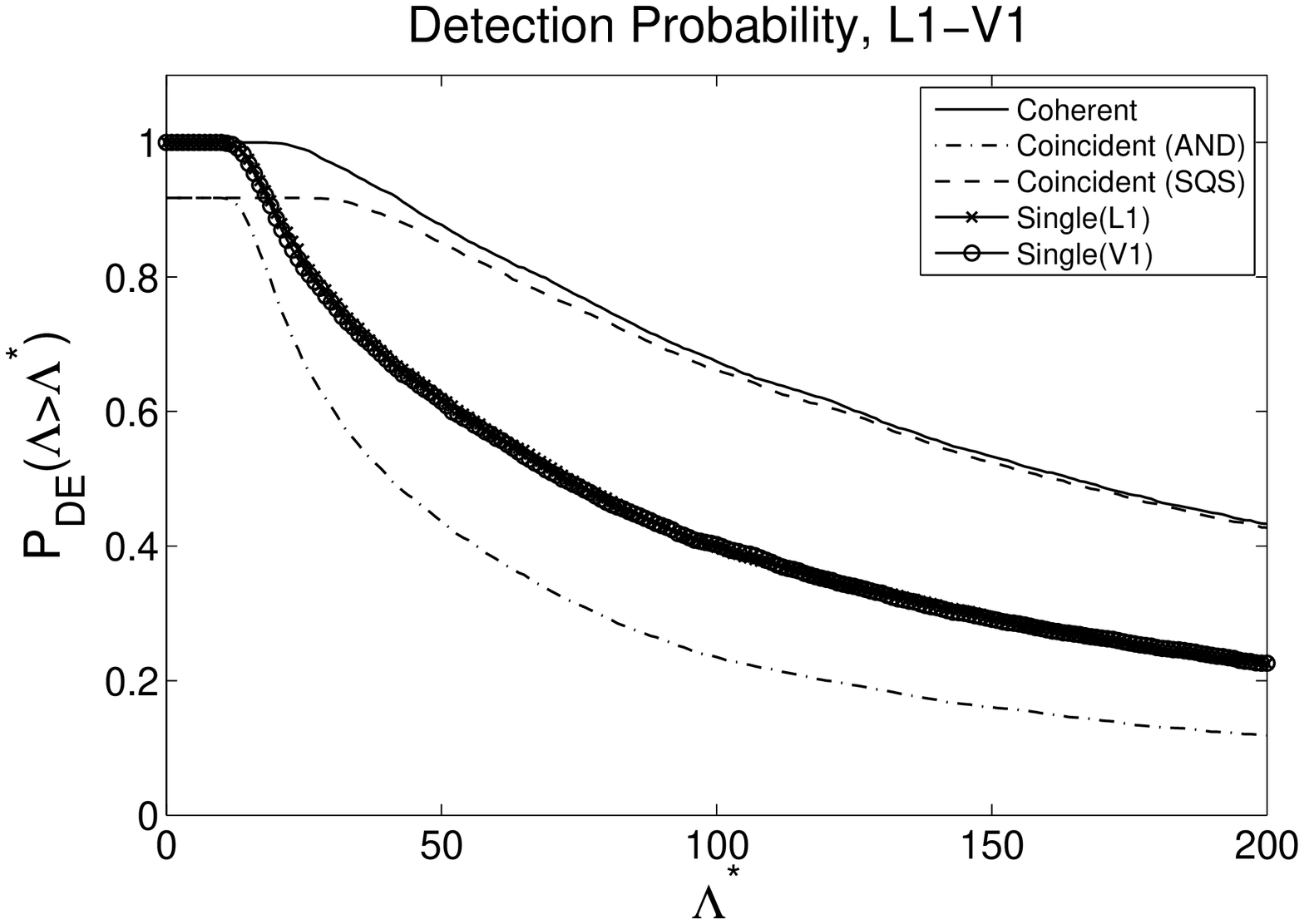} 
\mbox{(b) L1-H1}
\includegraphics[width=.5\textwidth]{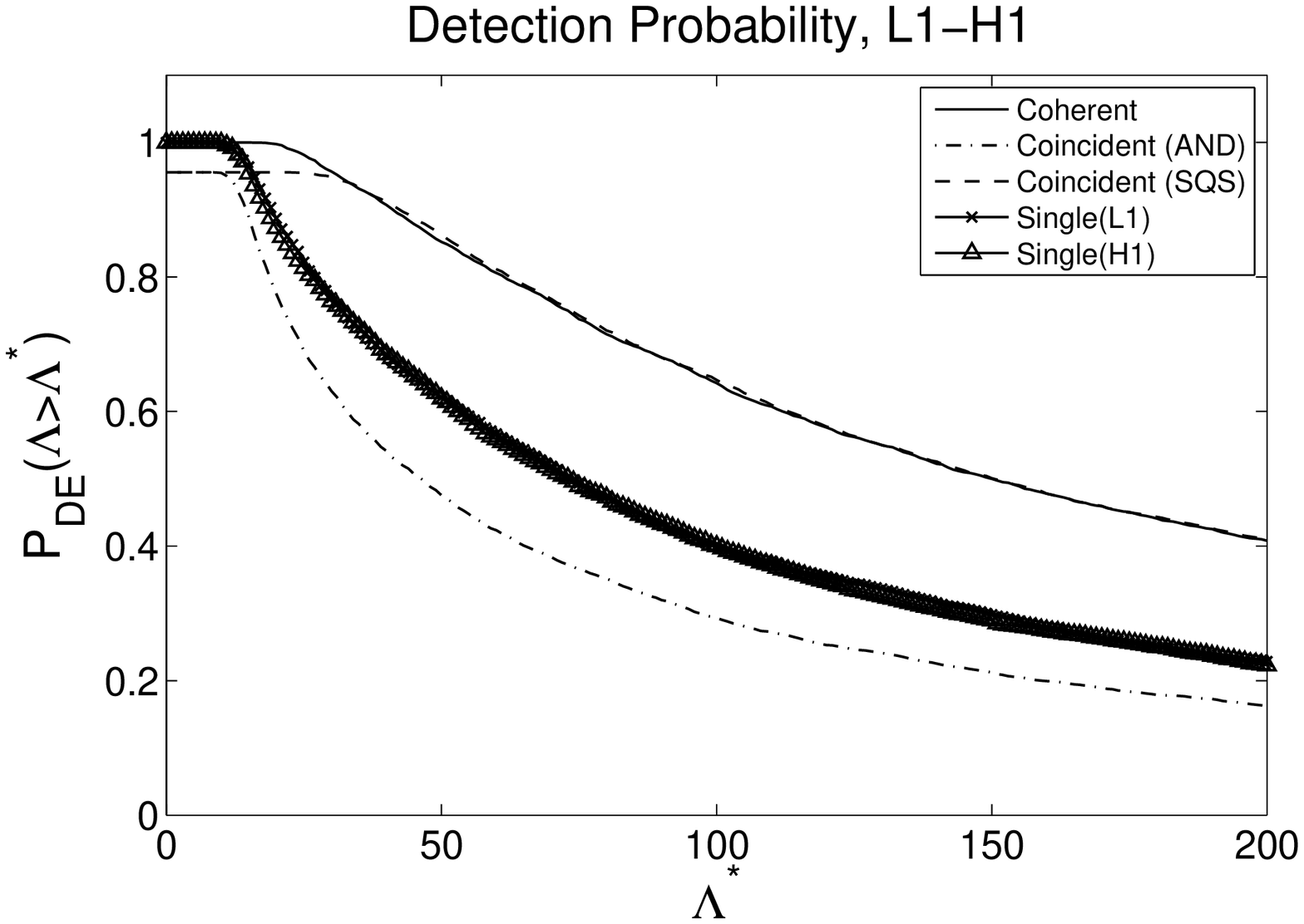} 
\mbox{(c) L1-K1}
\includegraphics[width=.5\textwidth]{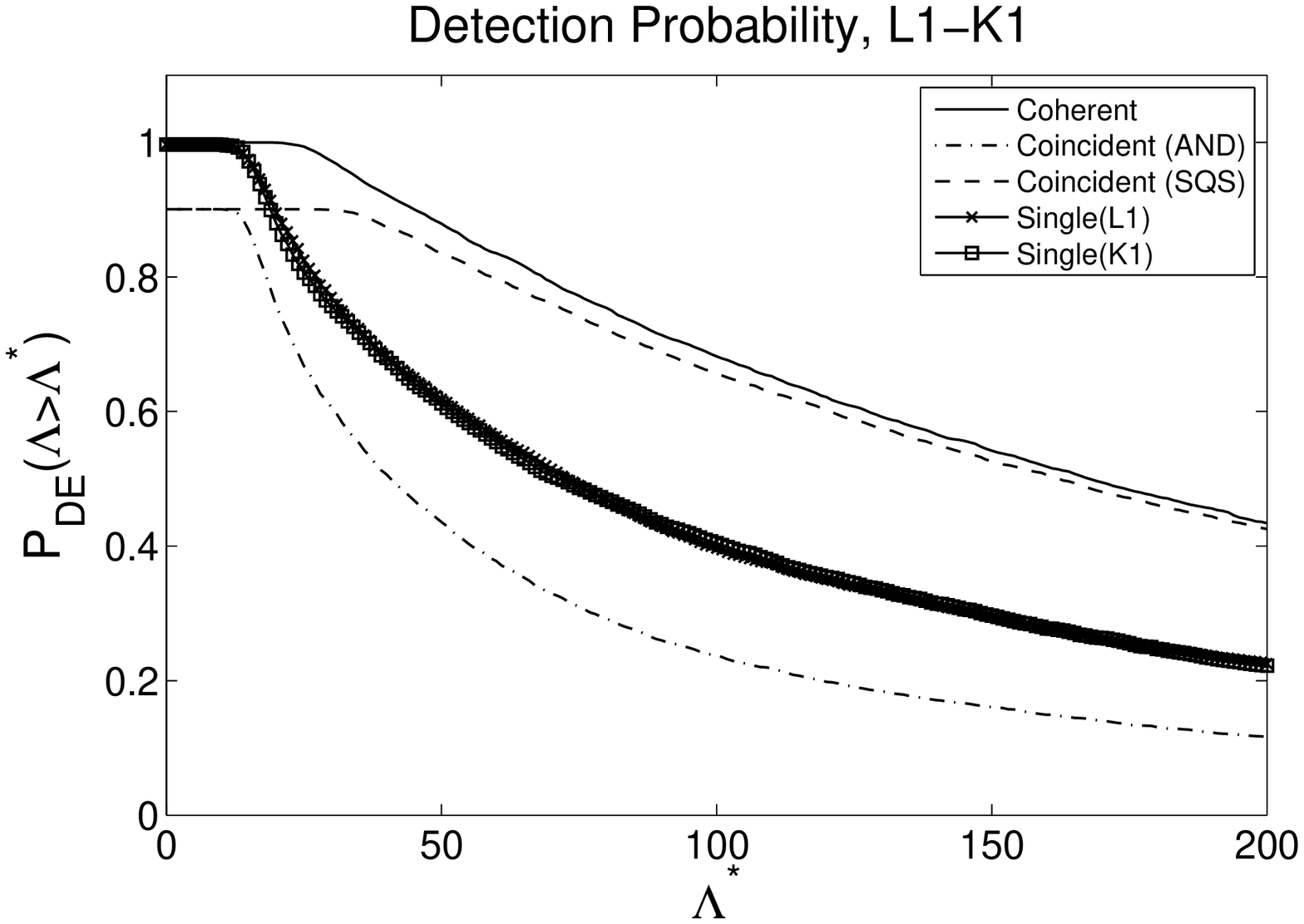} 
\mbox{(d) V1-K1}
\includegraphics[width=.5\textwidth]{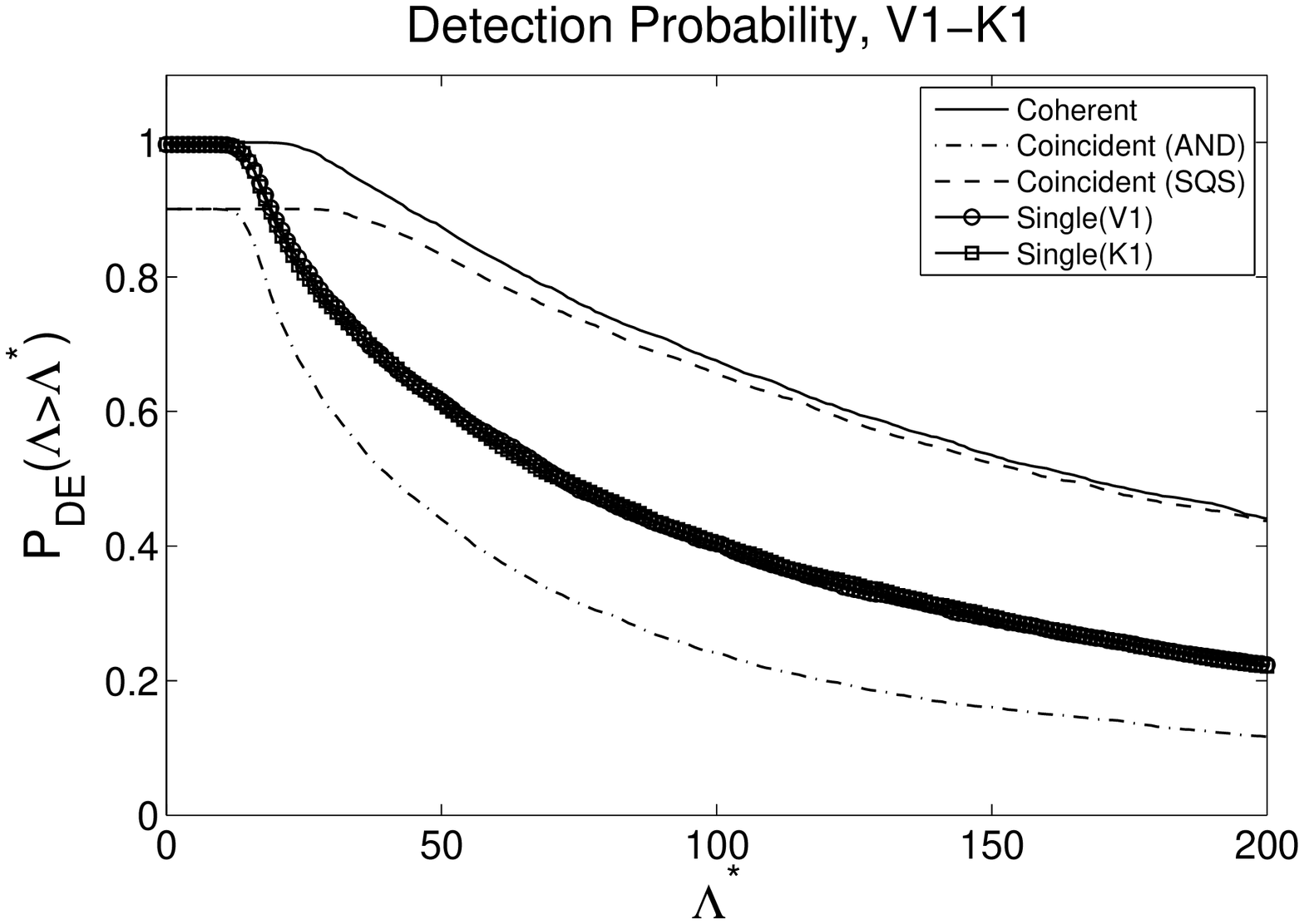} 
\caption{The detection probability for coherent, coincidence and single detector for 
distributed sources. Two single detectors' curves in each figures are almost identical and 
they can not be distinguished in the figures. 
}
\label{fig:DEdist}
\end{figure}

In order to obtain a fair comparison of the strategies, we consider a distribution of sources with different position, orientation and distance - we compare their average performance. The sources are taken to be uniformly distributed within 15 Mpc. The orientation and direction of the binary is randomly chosen from an uniform distribution (uniform in $\psi, \cos \iota, \phi$ and $\cos \theta$). The detection probability is shown in Fig. \ref{fig:DEdist}. 

\begin{figure}[htbp]
\centering
\mbox{(a) L1-V1}
\includegraphics[width=0.45\textwidth,height=0.22\textheight]{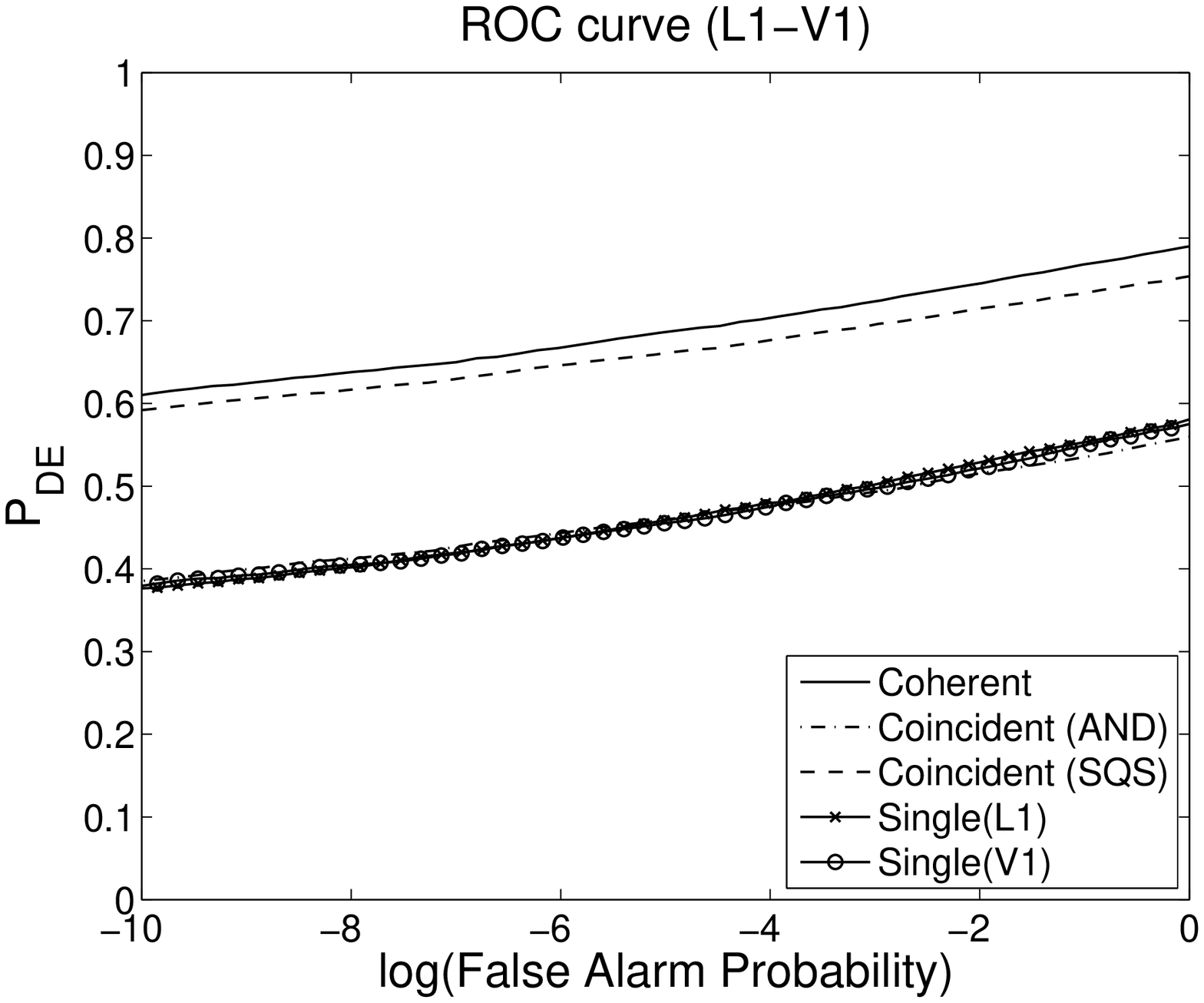} 
\mbox{(b) L1-H1}
\includegraphics[width=0.45\textwidth,height=0.22\textheight]{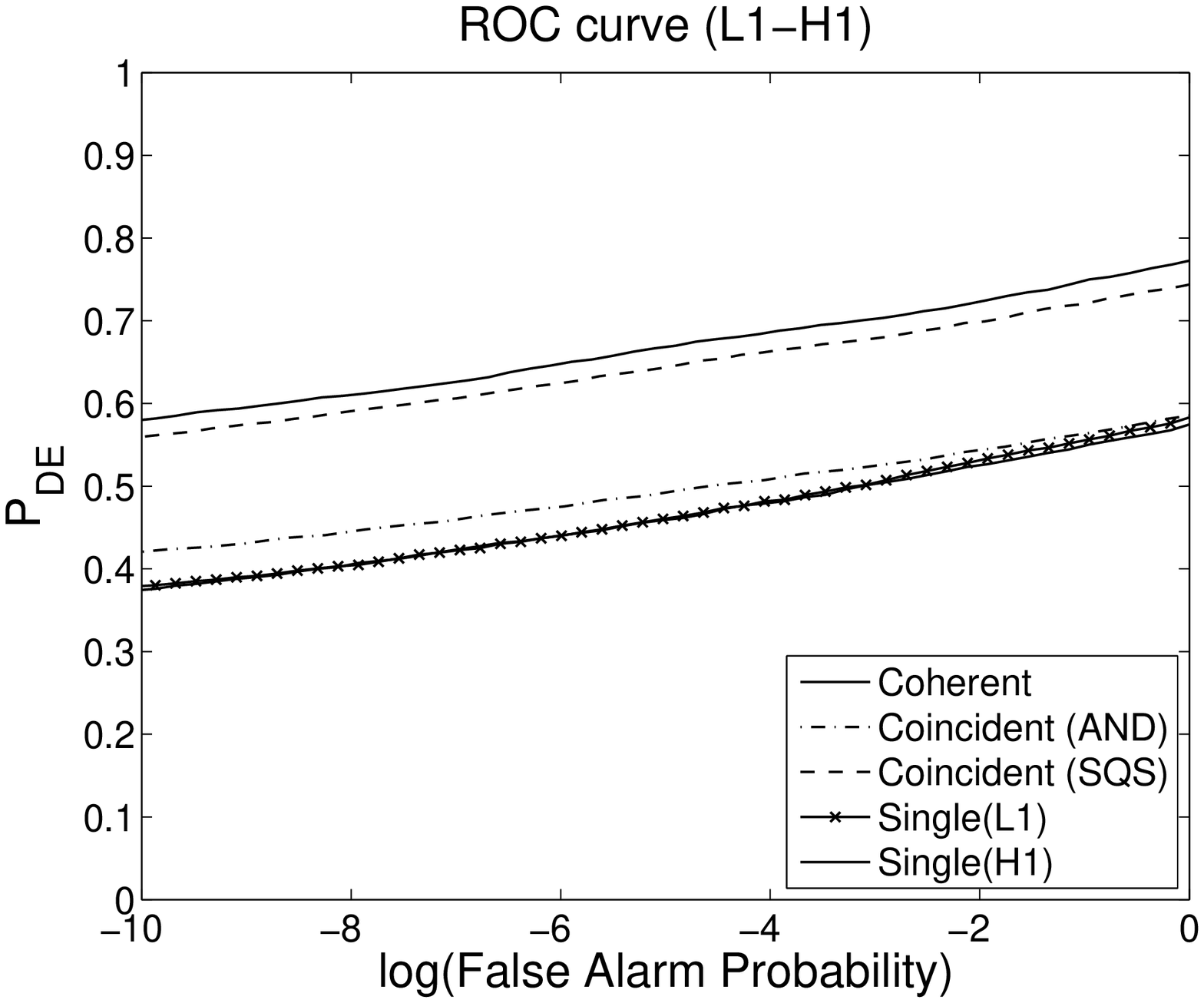} 
\mbox{(c) L1-K1}
\includegraphics[width=0.45\textwidth,height=0.22\textheight]{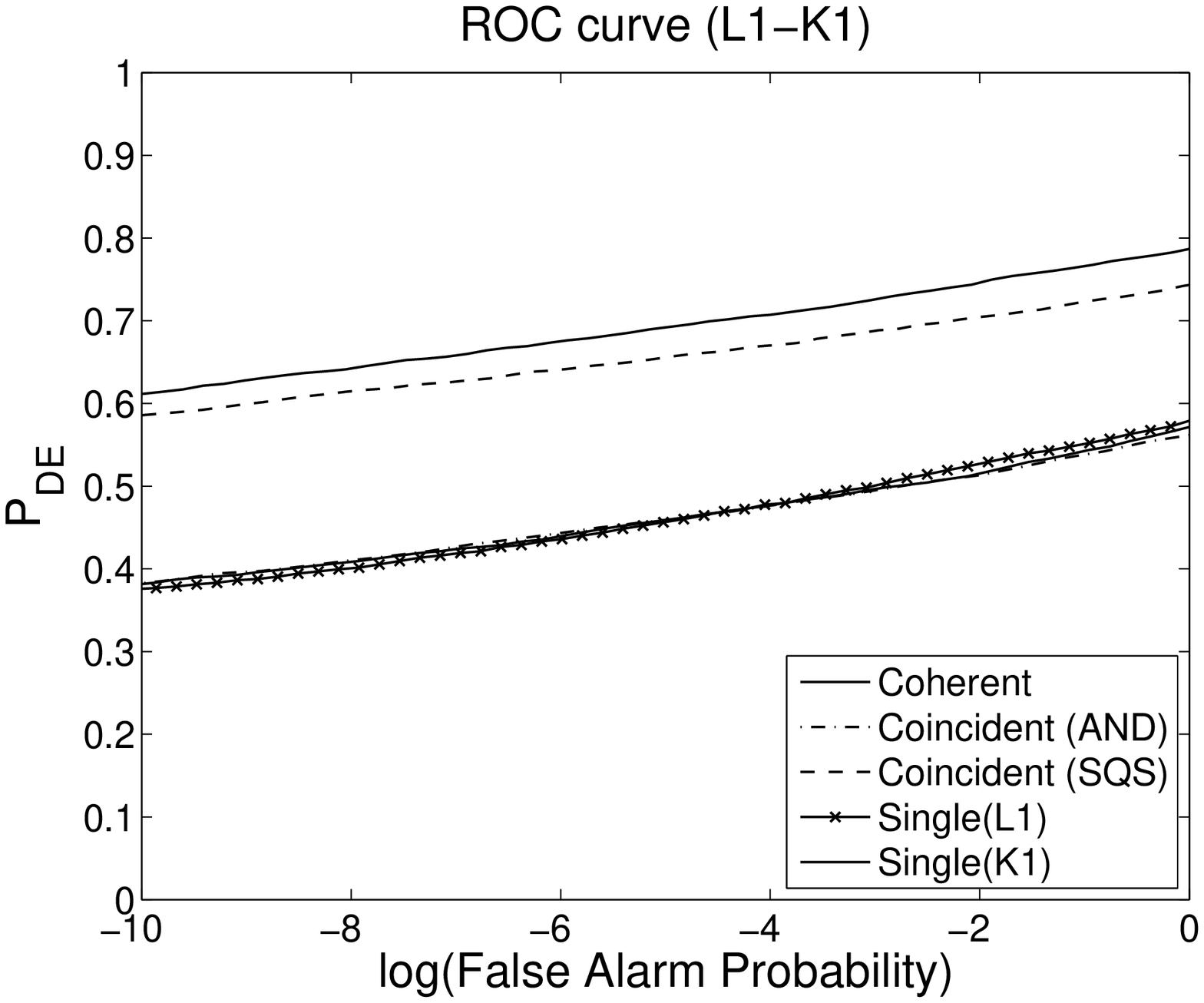} 
\mbox{(d) V1-K1}
\includegraphics[width=0.45\textwidth,height=0.22\textheight]{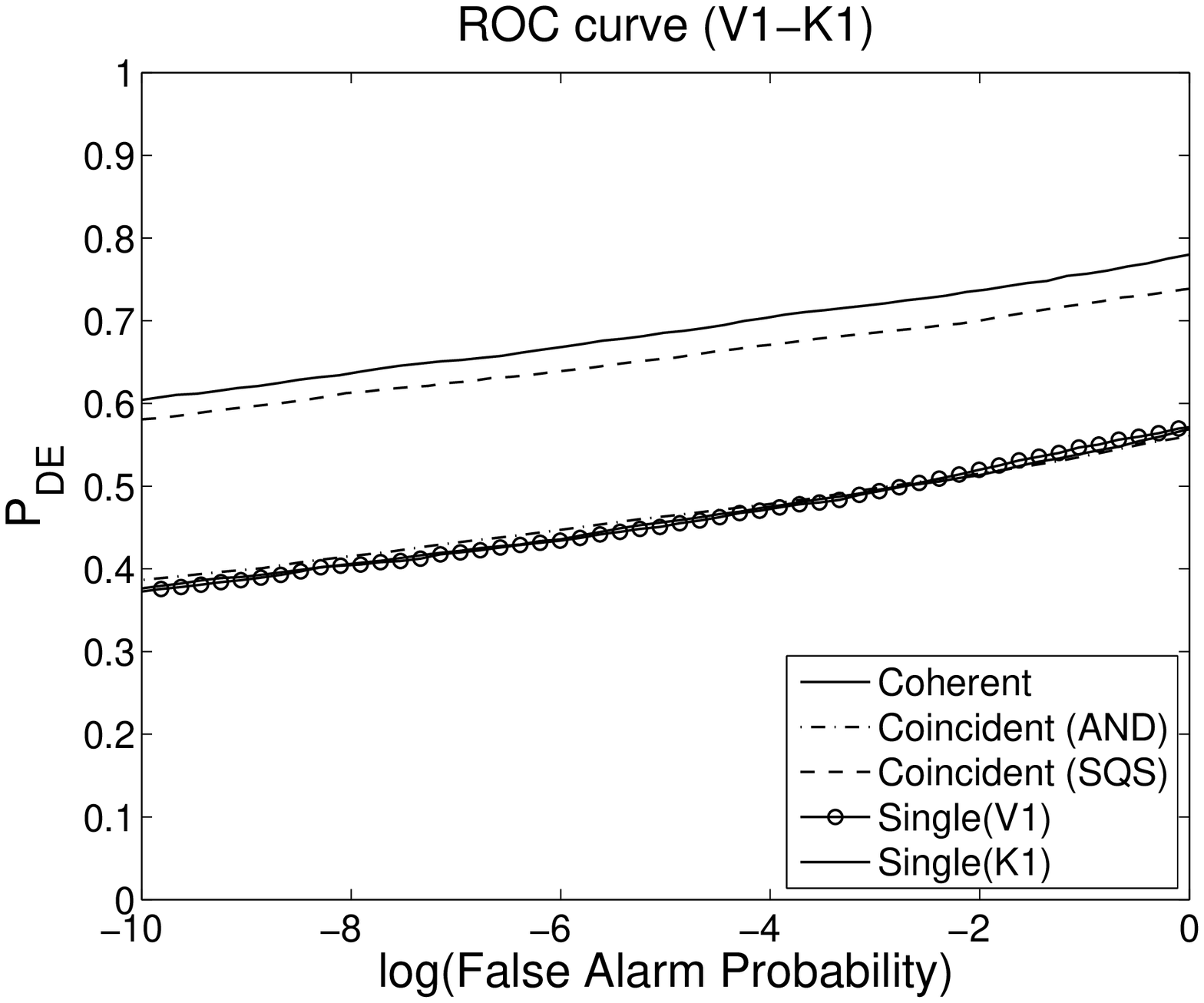} 
\caption{The ROC curves for coherent, naive coincidence and enhanced coincidence for (a) L1-V1, 
(b)  L1-H1, (c) L1-K1 and (d) V1-K1 for distributions of sources. } 
\label{fig:rocnet}
\end{figure}

We again consider a one year data train and mass ranges from 1$\msun$ to 40$\msun$. 
The ROC curves are drawn for L1-V1, L1-H1, L1-K1 and V1-K1 network in Fig.\ref{fig:rocnet}. 
A reduction of the factor of 5 for coincident strategies has been included in these plots. For the sake of comparison, single detector performance curves are also drawn. We have assumed that all the detectors have the noise PSD of initial LIGO. 

For example, in the L1-V1 case, at the false alarm probability of $10^{-5}$, the detection probability is 0.69 (coherent), 0.45 (naive coincidence), 0.66 (enhanced coincidence), and
0.45 (single). Thus, the coherent strategy performs around 50\% better than the naive coincident strategy. In paper-I for coaligned detectors at the same place, we concluded that the coherent detection is better than the coincident detection by about 25-40\%. Thus the difference in misaligned detectors is larger than the coaligned case. In the case of misaligned detectors, the amplitude of the signal in each detector differs, because of the different orientations of the  detectors. This difference produces smaller detection probability in the naive coincident strategy. This is manifest in Fig. \ref{fig:DEdist}.  Moreover, the detection probability of the naive coincident strategy is even less than each of the  single detector cases of L1 and V1. In the ROC plot (Fig. \ref{fig:rocnet}), however, the detection probability of the naive coincident strategy and that of the single detectors is nearly equal because the false alarm probability of the naive coincident strategy is smaller than the single detectors which compensates for the difference in the detection probability. The reason why the naive coincidence's curve and single detectors' curve in Fig. \ref{fig:rocnet} coincide is only by chance or coincidence. In the L1-H1 case, for example, the light travel time is smaller and the orientations of the detectors is nearly the same and therefore, the naive coincident strategy gives larger detection probability than the L1-V1 case for a given false alarm probability. 

Enhanced coincidence strategy performs better than the naive coincidence strategy, but we see that coherent strategy is still superior by around 5\%. Although the detection probabilities of the coherent and enhanced coincident strategies are nearly equal as can be seen in Fig. \ref{fig:DEdist}, the false alarm probabilities  of the  two strategies differ. This produces a difference in the ROC curves.

Note that although in the two coincident strategies, we reject the injected signal with $\Lambda_I\leq 16, I = 1, 2$ in the evaluation of the detection efficiency, we do not follow this procedure for the coherent strategy and in the single detector cases. Thus, if we take a threshold less than 16 in naive coincidence and 32 in enhanced coincidence, we cannot do a fair comparison. However, in the ROC curve of Fig.\ref{fig:rocnet}, the threshold range used in plotting the ROC curve is $35-70$ (naive coincidence) and more than 100 (enhanced coincidence). Thus, the rejection of the low amplitude signals in the evaluation of the detection efficiency does not affect the ROC curves. 

\section{conclusion}
\label{conclusion}

We compare the two strategies for analysis of network GW data for inspiraling binaries, namely the coherent and the coincident strategy. Analysis with real data will lead to actual comparison between the two strategies but as suggested by our findings under simplifying assumptions, the performance of the coherent strategy is  superior to the coincidence strategy on the whole. We distinguish two sub strategies in coincident detection, namely, naive coincidence and enhanced coincidence. For naive coincidence, the difference in performance, as compared with coherent, is even more conspicuous for misaligned detectors than coaligned detectors, for there are regions of the sky from where the signal may not be detected separately by the two detectors ruling out the possibility of coincidence detection, while, with a coherent detector, the source may still be visible. In fact this difference is so glaring that another strategy which we call enhanced coincidence needs to be devised.  The coherent strategy uses the statistic which is optimal in the maximum likelihood sense for the network. It inherently incorporates the information about the phase to decide on the detection for the statistic explicitly contains data from different detectors added with consistent phases. On the other hand coincident strategy treats the detectors separately, missing the crucial information about the phase altogether in consequence. This vital difference leads to superior performance of the coherent strategy. 

Although the coherent strategy is superior to coincident strategies,
the difference between the coherent and enhanced coincident strategies is not so large.
Only a relative improvement of about 5\% in the detection probability is obtained with the coherent strategy. One may ask whether there is any practical advantage in using the coherent strategy in the case of two misaligned detectors. Note however that the coherent method is not so computationally expensive compared with two coincident methods, since we do not take cross correlation of two detectors' data in the coherent strategy. Thus overall, the coherent strategy is a good detection method.

In future we would like to consider detector networks with more than two detectors. In case of coherent strategy, adding a detector to the network increases the sensitivity and therefore the detection efficiency of the coherent detector without exception. The locations in the sky for which the source is detectable also increases for coherent strategy upon addition of detectors thereby increasing the number of potential sources. In fact regardless of sensitivity of a detector in a network, adding a detector to the network, for coherent strategy, always improves the performance of the network. Despite the increase in computational cost, conducting coherent search will enhance the probability of detection of GW. The situation for enhanced coincidence is not so clear and needs investigation.

\begin{acknowledgments}
S. Dhurandhar acknowledges the DST and JSPS Indo-Japan international cooperative programme
for scientists and engineers for supporting visits to Osaka City University,
Japan and Osaka University, Japan.
H. Tagoshi and N. Kanda thank JSPS and DST under the same Indo-Japan programme
for their visit to IUCAA, Pune.
This work was supported in part by Monbu Kagakusho Grant-in-aid
for Scientific Research of Japan (Nos. 16540251 and 20540271).

H. Tagoshi, S. V. Dhurandhar and N. Kanda deeply regret the loss of their colleague and coauthor Himan Mukhopadhyay whose promising career was tragically ended by her untimely death during the course of this paper.

\end{acknowledgments}

\end{document}